\documentclass[aps,pra,twocolumn,superscriptaddress,floatfix, nofootinbib]{revtex4-1}
\usepackage{blindtext}
\usepackage{amsmath}

\usepackage{hyperref}
\hypersetup{
  colorlinks   = true, 
  urlcolor     = blue, 
  linkcolor    = blue, 
  citecolor   = blue 
}

\usepackage{bm}
\usepackage{gensymb}
\usepackage[normalem]{ulem}
\usepackage{euscript}
\usepackage[pdftex]{graphicx}
\usepackage{xspace}
\usepackage{xfrac}
\usepackage{enumerate}
\usepackage{enumitem}
\usepackage{braket}
\usepackage{float,fancyvrb}
\usepackage[T1]{fontenc}
\usepackage{lmodern}
\usepackage{lipsum} 
\graphicspath{{./Figures/}}
\usepackage{wrapfig}
\usepackage{scrextend}
\usepackage{comment}
\usepackage{url}
\usepackage{amssymb}
\usepackage{xcolor}

\definecolor{ao}{rgb}{0.0, 0.5, 0.0}

\setlist{nosep}

\begin{document}

\title{Multibody molecular docking on a quantum annealer}

\author{Mohit Pandey}
\thanks{These authors contributed equally to this work and are listed in alphabetical order.}
\affiliation{Menten AI, Inc., San Francisco, CA 94111, USA}
\author{Tristan Zaborniak}
\thanks{These authors contributed equally to this work and are listed in alphabetical order.}
\affiliation{Menten AI, Inc., San Francisco, CA 94111, USA}
\affiliation{Department of Computer Science, University of Victoria, Victoria, BC V8W 2Y2, Canada}
\author{Hans Melo}
\affiliation{Menten AI, Inc., San Francisco, CA 94111, USA}
\author{Alexey Galda}
\affiliation{Menten AI, Inc., San Francisco, CA 94111, USA}
\author{Vikram Khipple Mulligan}
\thanks{To whom correspondence should be addressed: vmulligan@flatironinstitute.org}
\affiliation{Center for Computational Biology, Flatiron Institute, 162 Fifth Avenue, New York, NY 10010, USA}

%\date{\today}

\begin{abstract}

Molecular docking, which aims to find the most stable interacting configuration of a set of molecules, is of critical importance to drug discovery. Although a considerable number of classical algorithms have been developed to carry out molecular docking, most focus on the limiting case of docking two molecules. Since the number of possible configurations of $N$ molecules is exponential in $N$, those exceptions which permit docking of more than two molecules scale poorly, requiring exponential resources to find high-quality solutions. Here, we introduce a one-hot encoded quadratic unconstrained binary optimization formulation (QUBO) of the multibody molecular docking problem, which is suitable for solution by quantum annealer.  Our approach involves a classical pre-computation of pairwise interactions, which scales only quadratically in the number of bodies while permitting well-vetted scoring functions like the Rosetta REF2015 energy function to be used.  In a second step, we use the quantum annealer to sample low-energy docked configurations efficiently, considering all possible docked configurations simultaneously through quantum superposition.  We show that we are able to minimize the time needed to find diverse low-energy docked configurations by tuning the strength of the penalty used to enforce the one-hot encoding, demonstrating a 3-4 fold improvement in solution quality and diversity over performance achieved with conventional penalty strengths.  By mapping the configurational search to a form compatible with current- and future-generation quantum annealers, this work provides an alternative means of solving multibody docking problems that may prove to have performance advantages for large problems, potentially circumventing the exponential scaling of classical approaches and permitting a much more efficient solution to a problem central to drug discovery and validation pipelines.

%Usually, the penalty strength of a one-hot encoding is chosen large enough to separate the energy bands comprising meaningful (physical) and non-meaningful (non-physical) states to avoid sampling non-physical states. However, we find that low-energy solution quality and diversity is reduced in this regime, compromising applications where sampling from excited low-energy states is useful, including drug discovery. Instead, we find that choosing a moderate value of the penalty strength, which promotes a high degree of intermixing between physical and non-physical states, ensures a larger diversity of low-energy solutions when sampled using the D-Wave quantum annealer, demonstrating a 3-4 fold improvement versus convention.

% Those exceptions which permit docking of more than two molecules scale poorly, given that the number of possible poses is exponential in the number of docking molecules, which essentially guarantees exponential use of either time or spatial resources if solution quality is to be maintained.

\end{abstract}

\maketitle

\section{Introduction}

The development of new drugs typically requires the screening of hundreds of thousands or millions of molecules spanning a large chemical space. In the last few decades, computer-aided drug design methods have attempted to accelerate the discovery of lead drug candidates \cite{sliwoski2014computational, mulligan2020emerging}. More recently, quantum computing has emerged as a promising tool that can complement classical computational methods in the drug discovery process \cite{cao2018potential, outeiral2021prospects, zinner2021quantum}. Quantum computers take advantage of the unique properties of quantum mechanical systems, such as superposition and entanglement, to carry out computations in a manner inaccessible to classical computers.  This permits new and possibly better-scaling means of solving optimization problems \cite{banchi2020molecular, mulligan2020designing, mato2022quantum, casares2022qfold}, performing quantum chemistry simulations  \cite{tilly2021variational, cao2019quantum}, or producing machine learning models \cite{batra2021quantum, li2021drug}.  There is therefore considerable interest in developing quantum equivalents of the poorest-scaling classical methods in computational drug discovery.

%Given that we are in the noisy intermediate-scale quantum era \cite{preskill2018quantum},  which is characterized by quantum hardware having a limited number of available qubits, low inter-qubit connectivity and imperfect qubit control, 

In the present noisy intermediate-scale quantum era \cite{preskill2018quantum}, it is believed that optimization might be one of the earliest applications to benefit from quantum resources \cite{Moll_2018}. For optimization problems relevant to drug discovery involving exponentially large search spaces, classical simulated annealing \cite{kirkpatrick1983optimization, vcerny1985thermodynamical} is one of the most-used approaches, and has shown great success in finding either optimal or near-optimal solutions. Though simulated annealing is a heuristic that surrenders guarantees of reaching the global optimum in all but the limit of an infinitely long annealing schedule \cite{geman1984stochastic}, it has empirically shown fast convergence to optimal or near-optimal solutions in a wide variety of cases \cite{suman2006survey, romeo1991theoretical}.  Until the recent advent of machine learning based methods \cite{jumper2021alphafold}, simulated annealing-based searches of protein conformational space represented the state of the art for protein structure prediction \cite{kallberg_template-based_2012,ovchinnikov_protein_2018}, and even now, simulated annealing-based exploration of chemical space remains the state of the art for amino acid sequence design, used in programs like the Rosetta software suite to create druglike peptides and proteins \cite{leaverfay2011rosetta,strauch_computational_2017,mulligan_computationally_2021,hosseinzadeh_anchor_2021}. However, while sampling the rugged energy landscapes of problems such as protein folding \cite{li2004accelerated, li2005parallel} and spin-glass models \cite{das2008colloquium}, simulated annealing can get stuck in local minima separated by high barriers.

%For optimization problems, classical computers usually search for the optimal solution in a large phase space by either exhaustive search or heuristic, iterative optimization algorithms, each of which require testing possible solutions one at a time [Note that this argument doesn't work for classical parallel computing]. Quantum computers, by contrast, represent all possible solutions as a quantum superposition, which allows quantum algorithms then to shift the probabilities of this superposition so that on measurement, the correct solution is returned with a high probability. This method is with the advantage of permitting a parallelism over every solution with an exponential savings in computational resources versus classical representations. [Maybe better to say -- Though simulated annealing, which is a classical heuristic algorithm, has shown great performance in wide variety of cases, it can take exponentially long time in sampling ground state for glassy problems (?). Quantum computing is expected to use quantum effects such as entanglement and tunneling to sample better but it's unclear there is any advantage. ]

Quantum annealing, a hardware-instantiated metaheuristic leveraging the quantum adiabatic theorem to find the global minimum of a given objective function, is a potentially powerful alternative to addressing such problems. Quantum annealing algorithms are expected to tunnel through ``high but narrow'' barriers in energy landscapes, but their potential advantages remain under debate \cite{ronnow2014defining, heim2015quantum,denchev2016computational}.  Though there have been efforts to study optimization on both quantum annealer and gate-based quantum hardware, there has not yet been clear demonstration of significantly-improved scaling for sampling the lowest-energy state on currently available hardware as compared to classical approaches. However, for drug discovery efforts, we are interested in finding a large set of distinct favourably-scoring solutions rather than a single most favourable solution, for two reasons.  First, optimization requires the use of a model scoring function, often based on experimental data, that can only approximate physical energies (\textit{e.g.}, Rosetta's REF2015 scoring function \cite{das2008macromolecular,kaufmann2010practically,alford2017rosetta,leman2020macromolecular}, OSPREY's energy functions \cite{hallen2018osprey}, or molecular dynamics force fields such as CHARMM \cite{patel_charmm_2004,huang_charmm36m:_2017} or AMBER \cite{wang_development_2004,maier_ff14sb:_2015}).  This means that the lowest-energy state under such a model might not correspond exactly to the true lowest-energy state, making the set of slightly higher-scoring states of interest.  Second, computational predictions in drug discovery are best taken as hypotheses to be tested rather than as reliable facts, and given that \textit{in vitro} and \textit{in vivo} experiments inevitably have some false negative rate themselves, it is useful to have a ranked pool of hypotheses (\textit{e.g.} designed amino acid sequences likely to perform a particular function, or predicted docked configurations likely to be maximally stable) to carry forward for experiments rather than a single prediction, in order to maximize chances of finding a hit.

Although the importance of both sequence diversity and molecular structure diversity for drug discovery has been long understood \cite{dean1999molecular, galloway2010diversity, gorse1999molecular, huggins2011rational}, only recently has the diversity of sampled solutions from quantum algorithms been studied. King \textit{et al.}, for example, examined solver performance with respect to diversity.  These authors defined diversity in terms of time-to-all-valleys, a metric that measures a solver's ability to sample at least once from all energy landscape wells containing a ground state \cite{king2019quantum}. This metric is specific to problems with many degenerate (\textit{i.e.} isoenergetic) ground states. On the other hand, Mohseni \textit{et al.} and Zucca \textit{et al.} studied diversity in terms of number of near-optimal solutions that have mutual Hamming distance larger than a certain threshold \cite{mohseni2021diversity,zucca2021diversity}. These studies each showed that the D-Wave quantum annealing processor can outperform state-of-the-art classical solvers on certain instances of synthetic problem classes in sampling diverse, low-energy solutions.

The present study offers two main contributions.  First, we map the multibody molecular docking problem to a quadratic unconstrained binary optimization (QUBO) formulation, using a one-hot solution encoding, for evaluation by a quantum annealer.  The multibody docking problem occurs repeatedly in drug discovery, both when predicting structures of oligomeric macromolecular targets, and when validating designed drugs by predicting their binding mode to their target (often with accompanying water molecules) \cite{meng2011molecular, xiao2018multi, van2021information}.  Due to the exponential scaling of the solution space with the number of bodies, a more efficient means of solving this problem would represent a major advance for computational drug design efforts.  Second, we provide insight into the ideal tuning of hyperparameters to permit the quantum annealer to sample optimal and diverse near-optimal solutions more efficiently. Specifically, we show that tuning the one-hot penalty strength to allow intermixing between meaningful (physical) and non-meaningful (non-physical) basis states can lead to increased sampling of low-energy and structurally diverse solutions with the D-Wave Advantage quantum annealer.

\subsection{Mapping multibody docking to a quantum annealer}

Here we consider the multibody molecular docking problem, both as a real-world problem of considerable biological interest, and as a case study for studying structural diversity of solutions of a real-world problem as sampled on a quantum annealer (the D-Wave Advantage system). This problem involves searching for the optimal spatial configurations (or ``poses'') of $N>2$ molecules in a solution space that scales exponentially with $N$. Often, this problem is considered at the level of two bodies, as in implicit-solvent protein-ligand docking \cite{combs2013small, deluca2015fully}. However, many biological complexes consist of more than two bodies: for example, hemoglobin is comprised of four independent protein subunits which closely associate in complex \cite{perutz_structure_1960}.  Indeed, a large fraction of proteins have quaternary structure, and nearly all exist at least transiently in complex with other molecules \cite{ali_protein_2005}. Even in the case of protein-ligand docking, water molecules can play a key role in protein-ligand recognition, so that this is natively a multibody docking problem \cite{xiao2018multi, roberts2008ligand}.

Recent classical computational methods \cite{rosell2020docking} such as HADDOCK \cite{karaca2010building}, ATTRACT  \cite{de2015web, de2012attract}, EROS-DOCK \cite{ruiz2020using} and MLSD (Multiple Ligands Simultaneous Docking) \cite{li2010multiple} attempt to address the importance of the multibody docking problem, but their performance scales poorly when a large number of bodies are involved. For example, HADDOCK can only carry out simultaneous multibody docking for up to six bodies \cite{karaca2010building}. To deal with the intensive computational resources required for simultaneous multibody docking in the large $N$ limit, HADDOCK instead docks $<N$ bodies in iterative fashion until all $N$ bodies are docked. However, this approach fails to consider the full $N$-body configuration space, and so is prone to missing globally-optimal configurations achievable only through simultaneous docking. Quantum parallelism, in contrast, allows simultaneous consideration of all possible configurations, offering potential advantage over classical methods.

In this work, we map the multibody molecular docking problem to a QUBO formulation, allowing sampling using the D-Wave quantum annealer. The quality of solutions produced by any multibody docking method of course depends not only on the sampling method but also the scoring function used.  Our approach works for any energy or scoring function that can be expressed as a sum of one-body internal energies and two-body interaction energies. For demonstration purposes, we use the Rosetta REF2015 energy function \cite{alford2017rosetta}, the excellent accuracy of which allows us to test our predicted docked configurations against crystallographic data. We pre-compute one- and two-body energies classically, an approach that scales favorably (quadratically) with the number of bodies, $N$, and reserve the quantum annealer for the search phase, which requires time that is exponential in the number of bodies $N$ when performed classically. This can be contrasted with the approach of Casares \textit{et al.} for predicting protein structure on a quantum computer, in which energies for all possible configurations were classically pre-computed, losing any possible advantage gained during the search phase \cite{casares2022qfold}. Our approach is also distinct from previous studies of molecular docking using quantum computers, which predominantly focused on two-body molecular docking.  Mato \textit{et al.} mapped molecular unfolding, which is one of the steps of two-body molecular docking, to a quantum annealer \cite{mato2022quantum}. Banchi \textit{et al.} mapped two-body molecular docking to finding the maximum weighted clique in a graph, which can be sampled on a Gaussian Boson Sampler \cite{banchi2020molecular}. Their mapping requires heuristically identifying pharmacophore points for a ligand-receptor pair (a coarsely-grained representation of the molecular geometry), while our approach considers all the atoms of the bodies involved in multibody docking.  Importantly, where approaches based on pharmacophore distance graphs inevitably have degenerate symmetry-related ground states, our approach respects the chirality of biomolecules and their complexes.

%Also, our molecular docking approach can always differentiate between chirality of molecules and more easily capture of internal degrees of freedom, unlike their method.  
%\tz{[We need to write a few lines comparing our work to previous quantum molecular docking papers. Note that ProteinQure paper can't deal with chirality of molecules and let's look whether they use any realistic energy score.]}

\subsection{Increasing structural diversity of sampled solutions on the D-Wave quantum annealer}

As a proof of concept for our method, we studied the diversity of low-energy states of a three helix bundle rigid docking problem. We used an all-atom representation of each of the three helices, pre-computing internal (one-body) energies and helix-helix interaction (two-body) energies with a custom-written Rosetta application.  Due to the limited number of qubits and inter-qubit connectivity available on D-Wave Advantage 4.1, we considered only translation in the configuration space, but our approach can be easily extended to include rotational transformations and internal degrees of freedom for each molecule. While sampling the QUBO formulation on D-Wave, we discretized the translation space, fixed one helix, and used a ``one-hot''  encoding to represent translational gridpoint indices for each mobile helix relative to the fixed helix. Though there have been efforts to find alternative encodings \cite{chen2021performance, chancellor2019domain}, one-hot encoding remains popular when sampling discrete integer (higher-than-binary) quadratic models on the D-Wave system \cite{kumar2018quantum, rieffel2015case, stollenwerk2019flight, lucas2014ising}. Usually, the one-hot penalty strength is chosen to be larger than the energy scale of the problem QUBO to ensure that low-energy states correspond to physically-meaningful states (\textit{i.e.}, bitstrings with one $1$ per register, assigning exactly one translational gridpoint for each body) while high-energy states correspond to non-physical states (\textit{i.e.} bitstrings assigning no gridpoint or multiple gridpoints to the same body). However, we find that choosing the penalty strength to be large reduces both solution quality and low-energy solution diversity sampled on the D-Wave annealer given the one-hot multibody molecular docking problems that we test. Instead, we find that choosing moderate values of the penalty strength, which promotes intermixing of physical and non-physical states, ensures both improved solution quality and the largest diversity of physically meaningful, low-energy solutions when sampling our one-hot encoded problems on the D-Wave system. %The specific moderate values depend upon the metric chosen (solution quality versus solution diversity).

%\tz{Do we need to justify here why we believe solution diversity and quality to be improved with a reduced gamma penalty? After all, this is one of the main results of our work... Also, note that the reasoning is slightly different between greedy and non-greedy search. For NG, D-Wave seems to preferentially sample excited states (NEED REFS), so we need to make sure the physica states are excited enough... For G, local minima are physical; reduce the number of minima by reducing gamma and you improve the chance of reaching low-energy minima.}

%\mpand{Great question. I suggest that we put our justification in the discussion section at the end rather than here. This intro section is already getting too long even without including all of novel results (\textit{e.g.} TTTD) }

%\vmullig{ Agreed -- in the discussion, not here. }

\subsection{New findings}

In sum: (1) We contribute a one-hot encoded QUBO formulation of the multibody molecular docking problem executable by current quantum annealing hardware, requiring a classical pre-computation that scales polynomially in the number of bodies and simulation space size. (2) Sampling our QUBO formulation of multibody molecular docking on the D-Wave Advantage 4.1 quantum annealer, we find that the lowest-energy predicted configuration, which by exhaustive enumeration corresponds to the ground state allowed by our model, has a root mean square deviation (RMSD) of 2.4 $\AA$ from the native crystallographic structure of a three helix bundle. (3) We find that choosing moderate values of the penalty strength parameter of our model improves the speed of sampling diverse low-energy solutions, including the ground state, by a factor of 3-4 times over the penalty strength of common practice.

\section{Methods}

We organize our methods as follows. In section \ref{QUBO formulation}, we present a QUBO formulation to the multibody molecular docking problem; in section \ref{Energy function}, we discuss our choice of energy function; in section \ref{Evaluation metrics}, we outline the measures by which we assess the performance of our quantum annealing and simulated annealing approaches; and in section \ref{Problem setup}, we present the specific multibody docking problem that we investigate. Sections \ref{Quantum annealing} and \ref{Simulated annealing} outline our approach to scanning over the hyperparameters of our quantum annealing protocol and a QUBO simulated annealing protocol to allow their comparison, and the experiments we execute to judge their performance.

\subsection{QUBO formulation}\label{QUBO formulation}

Quantum annealing employs the adiabatic theorem of quantum mechanics in an attempt to find the global minimum of a given objective function. Specifically, the theorem states that adiabatically changing the Hamiltonian of a quantum mechanical system in the ground state will not perturb the system from its ground state if there persists an energy gap that isolates the ground state from excited states \cite{born_beweis_1928}. Finding the global minimum of an objective function using this method then amounts to evolving the initial Hamiltonian and its ground state to a final Hamiltonian which encodes the objective function, and reading out the final ground state. In practice, however, finite annealing time, integrated control errors, readout error, and noise (thermal and quantum) force sampling of excited states as well.

The D-Wave quantum annealer used herein offers a programmable objective function in QUBO form, where the ground state of the initial Hamiltonian is an equal superposition over all possible solutions to the QUBO. Allowing $h(b_i)$ to be the one-qubit bias coeffient of the $i$-th qubit and $J(b_i, b_j)$ to be the two-qubit coupling coefficient of qubits $i$ and $j$, the QUBO function is then:

\begin{equation}
    H_{\text{QUBO}} = \sum_i h(b_i)b_i + \sum_{i>j} J(b_i, b_j)b_ib_j,
    \label{QUBO_Hamiltonian}
\end{equation}

\noindent where $b_i$ is a binary variable, \textit{i.e.}, $b_i \in \{0, 1\}$. The full set of $h(b_i)$ and $J(b_i, b_j)$ coefficients therefore define the problem to be solved, and the values that the associated qubits take on upon measurement represent a solution. We seek to encode the multibody docking problem according to this formulation.

%, and to compare the performance of multibody docking on a quantum annealer against QUBO and non-QUBO simulated annealing protocols, specifically investigating ground state energies and solution diversity

To start, we discretize the continuous, real space in which multibody molecular docking natively takes place to $T$ points. For our purposes, given the spatial resource limitations to our selected quantum annealing hardware (D-Wave Advantage 4.1), we will ignore internal and rotational degrees of freedom, though the approach is easily extended to include these degrees of freedom provided that sufficient qubits are available. Fixing one body to avoid configurational degeneracy due to translational symmetries, there exist $T^{N-1}$ possible unique configurations of $N$ bodies.

We enforce a one-hot encoding scheme, so that each of the $N-1$ mobile bodies is associated with one of $N-1$ registers of qubits of length $T$, each qubit corresponding to a specific point in our discretized space representing a particular translational offset of the given mobile body from the fixed body. We therefore require $(N-1)T$ qubits in total. Meaningful \textit{physical} states correspond to bit-strings which contain exactly one $1$ per register, as this places each body in exactly one location, while non-meaningful \textit{non-physical} states correspond to bit-strings other than those of physical states. In this QUBO formulation of the multibody molecular docking problem, the number of non-physical states $N_{\text{non-physical}}$ is exponentially more than number of physical states $N_{\text{physical}}$, \textit{i.e.}, $N_{\text{non-physical}}= \mathcal{O}(2^{(N-1)T}) N_{\text{physical}}$, where $T \gg 1$. One might consider it to be a qubit-wasteful strategy in having such a large number of non-physical states in the spectrum, but we later show its presence can allow us to tune the sampled diversity of low-energy physical states through intermixing between non-physical and physical states. We leave the comparison of our approach's performance with that of more qubit-efficient approaches for future work.

We designate body $N$ of the set of bodies $X=\{1, 2, ..., N\}$ to be the fixed body, so that the set of mobile bodies is then $X^{\prime}=\{1, 2, ..., N-1\}$. Given our one-hot formulation, the pairwise interaction between any $x\in X^{\prime}$ and body $N$ must be encoded in the $h(b_i)$ qubit bias terms of Equation \ref{QUBO_Hamiltonian}. Pairwise interactions between any two bodies in $X^{\prime}$ are encoded as $J(b_i, b_j)$ qubit coupling terms. These interactions between bodies are determined by pairwise free energy functions; we discuss these in section \ref{Energy function}.  (Note that in the present formulation, the one-body molecular energies are invariant with arrangement, and contribute only a constant offset to the energy of any given configuration.  Since they do not alter the relative energy or location of minima, they can be dropped from the problem, or used only when reporting solution energies.)

Whereas previously we identified qubits by a general index as $b_i$, we now track the body to which the variable applies explicitly in superscript and its position in subscript. Thus, for a mobile body $x$, its location at the $i$-th position within the space of $T$ points is represented as $b_i^x$, where $i\in {1,2, ..., T}$ and $x\in {1,2, ..., N-1}$. Now, to penalize the occurrence of non-physical solutions we apply a set of equality constraints to enforce that exactly one $b_i^x$ should be selected per mobile body $x\in X^{\prime}$. We begin with the constraint that $\sum_i b_i^x = 1$ per mobile body. To penalize the situation where each $b_i^x$ is returned as $0$, we then subtract $1$ and square both sides, so that our constraint on body $x$ is:

\begin{equation}
    \sum_{i>j}2b_i^xb_j^x-\sum_ib_i^x+1 = 0.    
\end{equation}

%\mpand{\sout{\noindent (Note that with $b_i^x \in \{0, 1\}$, $b_i^xb_i^x = b_i^x$.) }}

Adding these constraints to the objective function while ignoring constant offset and multiplying each by a tunable one-hot encoding enforcement penalty strength parameter, $\gamma$, yields the problem QUBO:

\begin{equation}
    \begin{split}
        H_{\text{MBD}} &= \sum_{x\in X^{\prime}}\sum_{i=1}^{T}\big(h(b_i^x)-\gamma\big)b_i^x\\
        &+\sum_{x\geq y}\sum_{i>j}(J(b_i^x, b_j^y) + \delta)b_i^xb_j^y.
    \end{split}
    \label{MBD_Hamiltonian}
\end{equation}

\noindent In the above, we define $\delta$ to be $2\gamma$ when $x=y$ and $0$ otherwise. With $N-1$ mobile bodies, the number of linear terms of the QUBO is $(N-1)T$, and the number of quadratic terms is ${(N-1)T\choose2}$, which scales as $\mathcal{O}(N^2T^2)$.

\subsection{Energy function}\label{Energy function}

We must supply a set of $h(b_i^x)$ and $J(b_i^x, b_j^y)$ terms to Equation \ref{MBD_Hamiltonian}, where, as noted previously, these terms correspond to classically computed pairwise interaction energies between bodies. This requires a classical pre-computation over each pair of bodies in the simulation space, so that each possible relative configuration between a pair of bodies is assigned an energy. The number of such pre-computations needed with one body fixed is exactly ${N-1\choose2}T^2 + (N-1)T$, which scales quadratically in the number of bodies. This compares well with respect to the size of the solution space, which scales exponentially in the number of bodies.

%\mpand{Highlight that solution is exponentially large so it's okay in doing this pre-computation}

Pairwise-decomposible energy functions are common in molecular modelling \cite{wang_development_2004, maier_ff14sb:_2015}.  Such energy functions are often sums of terms that can include van der Waals interactions, Lazaridis-Karplus solvation, and Coulombic electrostatic potentials, among others. If the bodies to be docked are peptides or proteins, additional contributions might include backbone-backbone, sidechain-sidechain, and sidechain-backbone hydrogen bond energies, Ramachandran preferences, and proline ring closure energies \cite{alford2017rosetta}. We highlight that as long as the energy function allows pairwise decomposable precomputation between bodies as described above, it is compatible with Equation \ref{MBD_Hamiltonian}.

For our purposes, we opt for the smoothed Rosetta all-atom energy function (REF2015\_soft), which was parameterized both by tuning parameters to fit protein X-ray crystal structure data, and to reproduce bulk properties of liquids in small-molecule molecular dynamics simulations \cite{park_simultaneous_2016,alford2017rosetta}. We use a custom-written Rosetta application to output to a file a list of the pairwise energies between bodies for all possible relative offsets in the simulation space.

\subsection{Evaluation metrics}\label{Evaluation metrics}

Usually, quantum annealing applications are evaluated with respect to how well they reach the ground state of the objective function that they encode. The time-to-solution (TTS) metric is often employed to this end, which measures the time it takes for the quantum annealing protocol to reach the ground state with a certain degree of confidence. In the case of multibody molecular docking, this ground state is the most stable configuration possible between the involved bodies, given the energy function.

While we are interested in the ground state and TTS, in the case of molecular docking, it is useful to understand as well those configurations which do not correspond to the ground state but are nevertheless of low energy. The number of such configurations for a given problem is its target diversity, with the corresponding performance metric, time-to-target diversity (TTTD), measuring the time it takes for the quantum annealing protocol to reach this target solution diversity. In the case of molecular docking, the threshold that separates high-energy solutions from low might reasonably taken to be zero, as this would typically indicate a configuration that is better than having all bodies at infinite separation (yielding zero energy).  This is a consequence of energy functions typically used in molecular modelling having Lennard-Jones-like two-body terms, which approach zero as the separation between bodies increases.

Given this, we set our target diversity to those physical solutions which exhibit negative energies, after correcting for inclusion of the one-hot penalty strength.  We evaluate both our quantum annealing and simulated annealing approaches by time-to-solution (TTS) and time-to-target diversity (TTTD) using the definition of diversity described above.

%\mpand{It should be noted that diversity of states in terms of unique low-energy states might actually correspond to similar conformations of biomolecules involved in molecular docking.}

It should be noted that if the simulation space is finely sampled with respect to the scale of the problem, several solutions with unique low energies might correspond to nearly the same physical configuration, so that all might be considered to encode the same configuration. For example, one body might be shifted by a tiny fraction of an {\AA}ngstrom in a certain direction, while not strongly affecting the overall relative arrangement of the bodies. Diversity is then measured over groupings of states that encode the same configuration but have slightly different energies. However, due to the limited number of qubits and low qubit connectivity of the D-Wave Advantage 4.1 quantum annealer, our simulation space is coarsely sampled. We checked that each unique low-energy state within our chosen simulation space corresponds to a distinct configuration by computing the RMSD with respect to the three helix bundle's native crystallographic structure.

The above discussion shows that we need to be careful about choosing the space in which distinction of states is measured. Zucca \textit{et al.} \cite{zucca2021diversity} measured diversity of low-energy states in terms of pair-wise Hamming distance between spin states. This metric might capture distinct states in spin models like Ising chain; however, while solving biologically-relevant problems via one-hot QUBO mapping, a set of spin states with similar Hamming distance might correspond to distinct physical configurations. We therefore consider distinction of states within the native space of the problem. %For example, imagine for our three-helix problem, there are two translation grid points. Then the set of $\{ (1010,1001), (1010,0110), (1001,0101),(0101,0110) \}$ have each pairwise Hamming distance of two but they correspond to distinct physical configurations.} %\textcolor{red}{[Check if it makes sense. For four-body problem, I feel this point is clearer. In that case, $\{(101010,101001), (101010,011010)] \}$ have Hamming distance of two but have very distinct physical configuration]} 

%Diversity is then measured over groupings of solutions that encode the same configuration. On the other hand, if the simulation space is coarsely-grained with respect to the scale of the problem, each unique low-energy configuration may be taken as a unique solution. This works for us perfectly as due to limited number of qubits available on D-Wave forces us to have a coarse-grained simulation space. Moreover, note that we are interested in the structural diversity in configuration space. }

%If the simulation space is finely-grained with respect to the scale of the problem, several solutions with unique low energies might correspond to nearly the same physical configuration, so that all might be considered to encode the same configuration. For example, one body might be shifted by a tiny fraction of an {\AA}ngstrom in a certain direction, while not strongly affecting the overall relative arrangement of the bodies. Diversity is then measured over groupings of solutions that encode the same configuration. On the other hand, if the simulation space is coarsely-grained with respect to the scale of the problem, each unique low-energy configuration may be taken as a unique solution.

%\mpand{We should explain here why we can't borrow Zucca et al's diversity for our studies?}

%\tz{We already do, in a sense. The "Now, it should be noted that if the simulation space..." paragraph goes into why we can treat each low-energy solution as unique. The only difference between our study and Zucca is that they group solutions.}

\subsection{Problem setup}\label{Problem setup}

\begin{figure}[ht]
\begin{center}
    
    \includegraphics[width=9cm]{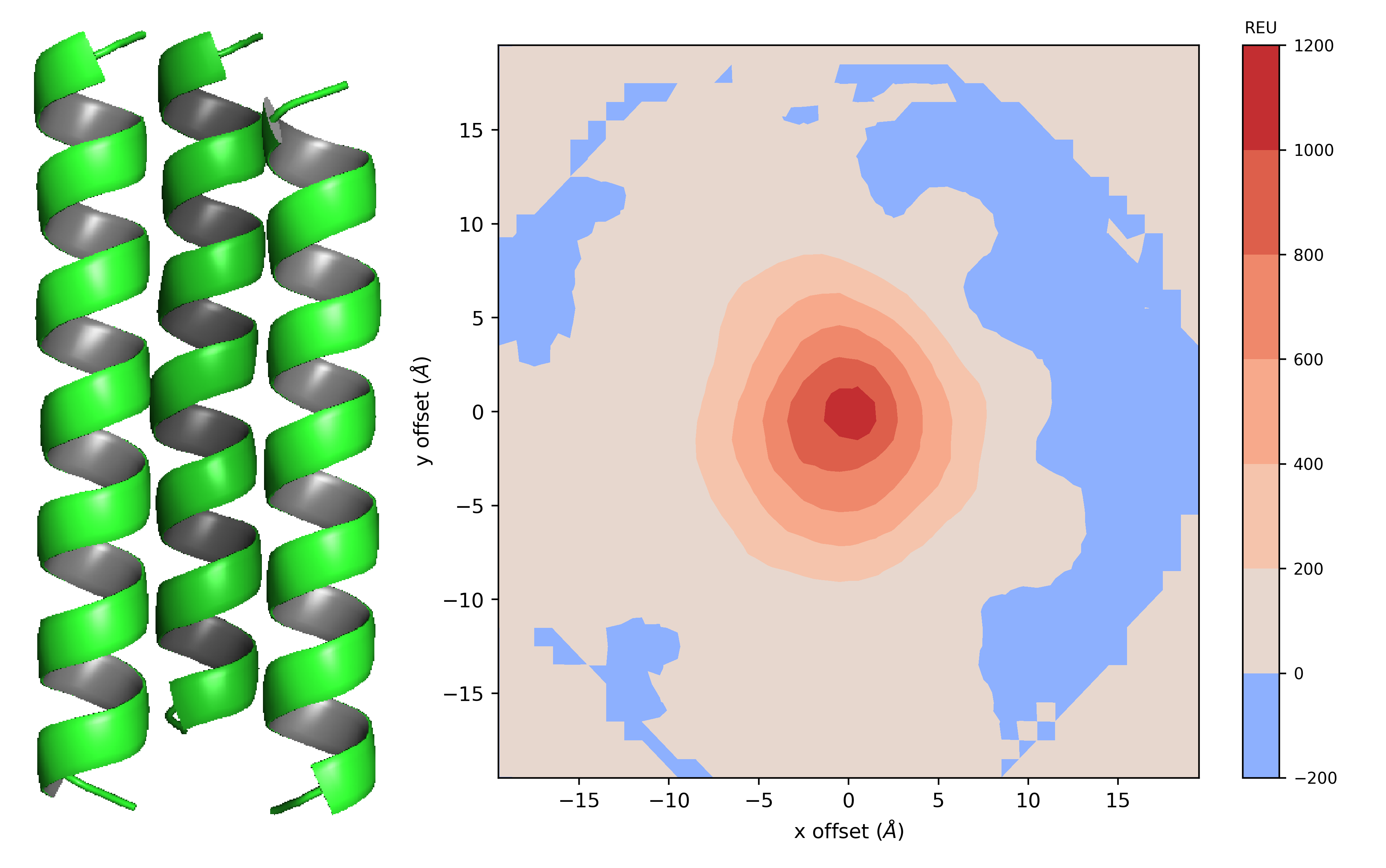}
    \caption{We tested our algorithm against a known three helix bundle (PDB ID: 4TQL, left panel). Loops were excised and helices were shortened to allow their free movement for our proof-of-concept study. The right panel shows the energy landscape of a pair of $\alpha$-helices selected from the bundle in two dimensions while fixing their internal conformations, rotational orientations, and z-offset (along the axis of the helices) to zero with respect to the native structure. For relative displacements less than the diameter of a single $\alpha$-helix ($\sim6$ $\AA$) the interaction energy rapidly grows large, while for relative displacements greater than $\sim15$ $\AA$, the interaction energy goes to zero. In between, when the helices are in close proximity without backbone clashes, sidechain-sidechain steric repulsion often results in high energies.  At slightly greater separation, there appear negative energy wells (blue) indicating mutual attraction. Note that though we are showing a two-dimensional energy landscape here, three helix rigid body docking in two-dimensional translation space requires us to sample configurations from a six-dimensional phase space.}
    \label{fig:three_helix_bundle}
\end{center}
\end{figure}

To test our method, we chose a computationally-designed three helix bundle whose structure has been experimentally confirmed by X-ray diffraction (PDB ID: 4TQL) \cite{huang_high_2014}.  This is shown in Figure \ref{fig:three_helix_bundle}. Although this protein is natively a single polypeptide chain, with three $\alpha$-helices separated by short loops, we excised these loops and shortened the helices to generate a three body problem, arbitrarily choosing one $\alpha$-helix to remain fixed in the simulation space.

%\tz{Why did we choose this over, say, hemoglobin or the ribosome, which don't require the pre-processing step of snipping looped regions?}
%\mpand{Because it's one of the simplest structure}

%Note that we performed docking of the three rigid helices, allowing no internal conformational variability, in a 2-dimensional translational space -- a subspace of the 6-dimensional space of free translations and rotations.
Next, we considered the size of simulation space which, given the two mobile bodies of our problem, can fit onto the Pegasus graph topology of the D-Wave Advantage 4.1 quantum annealer, the largest available quantum annealer at the time of experiment. Considering that with a one-hot encoding our approach requires a fully-connected QUBO graph, the connectivity and size of the available hardware limits embeddings to graphs containing up to 177 binary variables, admitting a largest chain length of 17 qubits \cite{d_wave_white_paper_advantage}.  A cubic simulation space with $N = 3$ bodies would therefore be limited to $T = 4\times4\times4$ points with the current hardware, the number of variables required being $(N-1)T$. To allow more points per dimension, we opted instead to simulate over a square planar space, perpendicular to the known axis of the bundle. With $N=3$ bodies, a maximum of $T=8\times8$ points may be simulated for such a space with the current hardware. To guarantee embedding of the problem graph and reduce the required chain length, we selected a slightly smaller space of $T=7\times7$. Physically, the separation of adjacent points within this space corresponds to a distance of 3 $\AA$, so that its physical dimensions are $18\times18$ $\AA$. This selection accommodates both the possibility of separation between the helices and their close proximity, given that the step size of 3 $\AA$ is smaller than the $\sim6$ $\AA$ diameter of an $\alpha$-helix. 

%We also tested over a $T=5\times5$ grid, with a separation distance between adjacent gridpoints of 4.5 $\AA$. This version of the problem maintains the physical dimensions of the simulation space at $18\times18$ $\AA$.

Using Rosetta, we then pre-computed the possible pairwise energies between each pair of rigid-body helices within each simulation space using REF2015\_soft, these energies forming the basis of the energy landscape for both the quantum annealing approach and simulated annealing approaches. For a given pair of $\alpha$-helices, their pairwise energy landscape is strongly positive when their separation distance is overlapping, goes to zero when their separation distance is far, and in between exhibits dips where mutual attraction occurs (see Figure \ref{fig:three_helix_bundle}). Since this small example problem allows only $49^2=2,401$ possible physical solutions, we exhaustively enumerated the energies of each of these solutions to provide a ground truth to compare to, specifically considering those physical states exhibiting negative energy. We shall refer to the full set of these states, constituting our target diversity, as $S$.

Note that with a problem size of $N=3$ and the grid size tested, the number of pairwise pre-computations is greater than the number of possible physical solutions. However, for $N>3$, the number of pairwise pre-computations grows only quadratically while the number of physical solutions grows exponentially with $N$.  This means that the number of pairwise pre-computations becomes exponentially less in $N$ than the number of possible physical solutions.

\subsection{Quantum annealing}\label{Quantum annealing}

After pre-computing the $h(b_i^x)$ and $J(b_i^x, b_j^y)$ energies for the three helix bundle problem, we capped highly-positive pairwise energy values at approximately twice the magnitude of the largest negative pairwise energy value. This reduces the variance of spin couplings and biases upon embedding the problem on the D-Wave QPU, which has limited dynamic ranges to its biases and couplers, thereby increasing the energy separation between lower-energy states upon embedding. The importance of this step is in reducing the impact of integrated control errors, which affect solution quality increasingly as separation between low-energy states decreases \cite{d_wave_ice}. Originally, the largest positive pairwise energy in the QUBO registered above $5000$, but with energy-capping based on the largest negative pairwise energy of $-38$, large positive energies reduced to $100$.
%\mpand{This reduces the variance of spin couplings and biases upon embedding the problem on the D-Wave QPU, which has limited dynamic ranges due to integrated control errors \cite{d_wave_ice}}. 

In keeping with the findings of Zucca \textit{et al.}, where a search over anneal time was performed and evaluated with respect to maximizing solution diversity over three classes of Ising problem \cite{zucca2021diversity}, we selected to work with an anneal time of 1 $\mu$s. We then conducted a preliminary search over the strength of the one-hot penalty constant $\gamma$ between $10^0$ and $10^3$, executing 25 replicates of $10^4$ successive samples each, recording the best energy and proportion of target solution diversity found per replicate. Further, we downsampled each replicate using a sliding window between $10^2$ and $10^4$ samples to investigate trends in best energy and solution diversity as functions of sample size. (Note that a metric computed using a smaller sample size gets averaged over a larger number of replicates.) For each solution returned by the D-Wave solver, we also applied a classical steepest-descent bit-flip optimizer (greedy post-processing) to find the local minimum in each solution neighborhood before recording solution diversity and quality as functions of one-hot penalty strength and number of samples as before. This optimizer has a run-time complexity of $\mathcal{O}(NT)$ per step, with experiments showing an average of 5 steps required per solution returned by the quantum annealer, irrespective of $\gamma$.

%\mpand{What's the greedy algorithm's run-time scaling in terms of a graph G with vertices V? Look at D-Wave doc}

Next, we computed from these data the success probability of finding the target diversity and ground state energy as a function of number of samples by downsampling, per each tested value of gamma. Specifically, we compute success probability by counting the number of replicates exhibiting target diversity or containing the ground state and dividing by the total number of replicates for a given value of one-hot penalty strength and given number of samples. For each one-hot penalty strength, we then fitted sigmoid curves of the form:

\begin{equation}
    y = \frac{1}{1+e^{\alpha(t-t_0)}},
\end{equation}

\noindent to the corresponding success probability functions, and took the resulting, fitted $t_0$ parameters as the number of samples required to reach target diversity or best energy with 50\% confidence.

\subsection{Simulated annealing}\label{Simulated annealing}

To understand better the performance and behaviour of the D-Wave quantum annealer, we also tested a classical QUBO simulated annealing solver. This approach works over exactly the same energy landscape as the quantum annealer, and also includes both physical and non-physical states whose separation is tunable by the strength of the one-hot penalty.  We specifically employ the D-Wave simulated annealing sampler, which accepts QUBO problems and updates bits according to the Metropolis-Hastings criterion. Using default parameters, we repeated the same experiments and analysis as for the quantum annealer.  %\mpand{Should we say that simulated annealing in the configuration space is fastest? It's speed is orders of magnitude faster}

\section{Results and Discussion}

We organize our results and discussion as follows. In section \ref{R&D:demonstration}, we provide a demonstration of a multibody docking problem solved in all-atom detail, using a robust Rosetta REF2015\_soft energy function, on the quantum annealer.  This serves as a proof of principle that multibody docking problems can be mapped to quantum hardware, and that a quantum annealer can produce meaningful results. In section \ref{R&D:mixing}, we discuss general effects of tuning the one-hot QUBO penalty strength on the intermixing between physical and non-physical states, and on the ruggedness of the energy landscape. Further, we point to a means of estimating the degree of this mixing and ruggedness in advance of sampling. In section \ref{R&D:diversity}, we present and discuss the specific dependence of solution diversity of the three helix bundle problem on the one-hot penalty strength, as sampled by D-Wave Advantage 4.1 and the QUBO simulated annealing solver, including time-to-target diversity results. In section \ref{R&D:quality}, we present and discuss the specific dependence of solution quality of our problem on one-hot penalty strength, including time-to-target solution results.

\subsection{Multibody docking on a quantum annealer proof of principle}\label{R&D:demonstration}

\begin{figure*}
    \includegraphics[width=5.5cm]{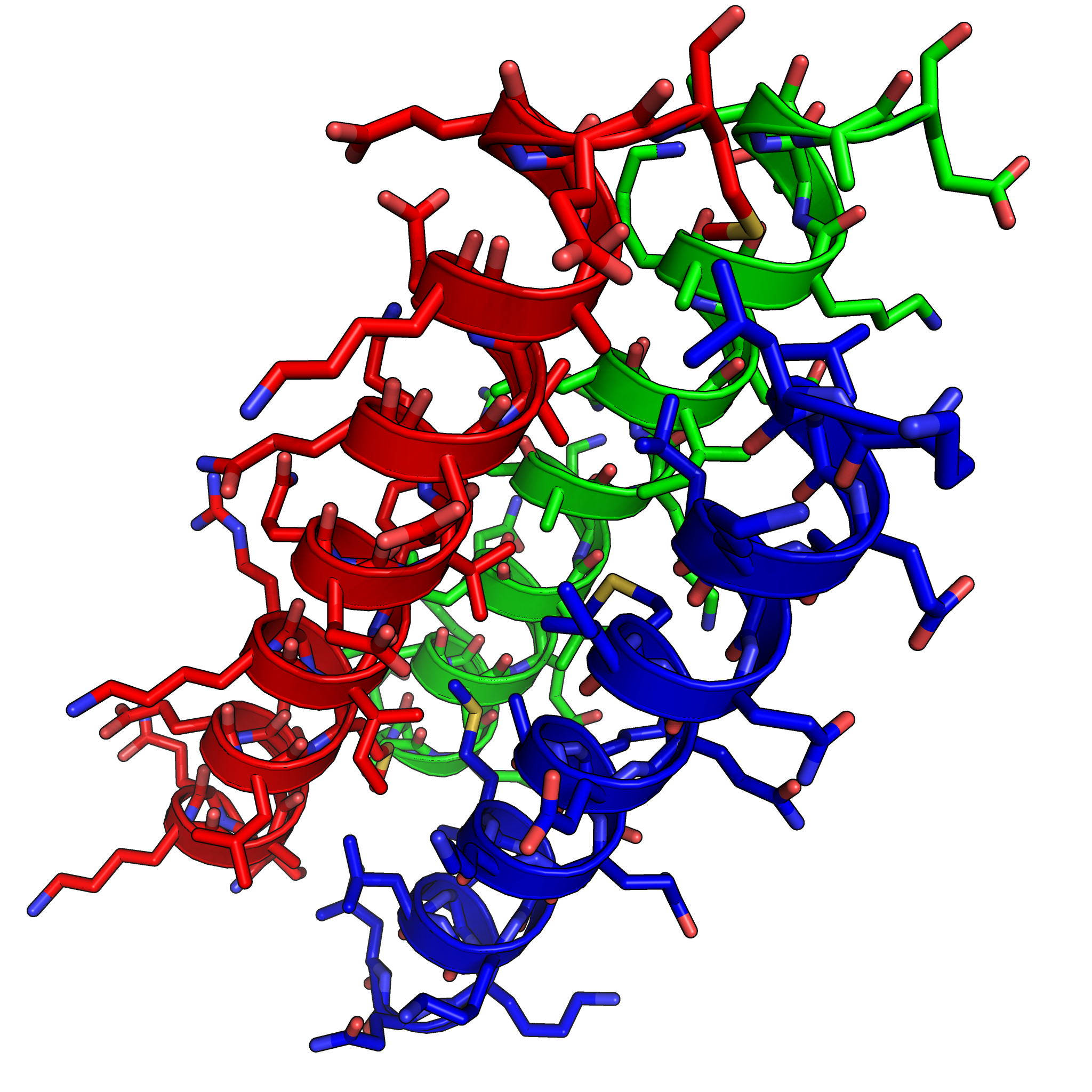}
 \includegraphics[width=5.5cm]{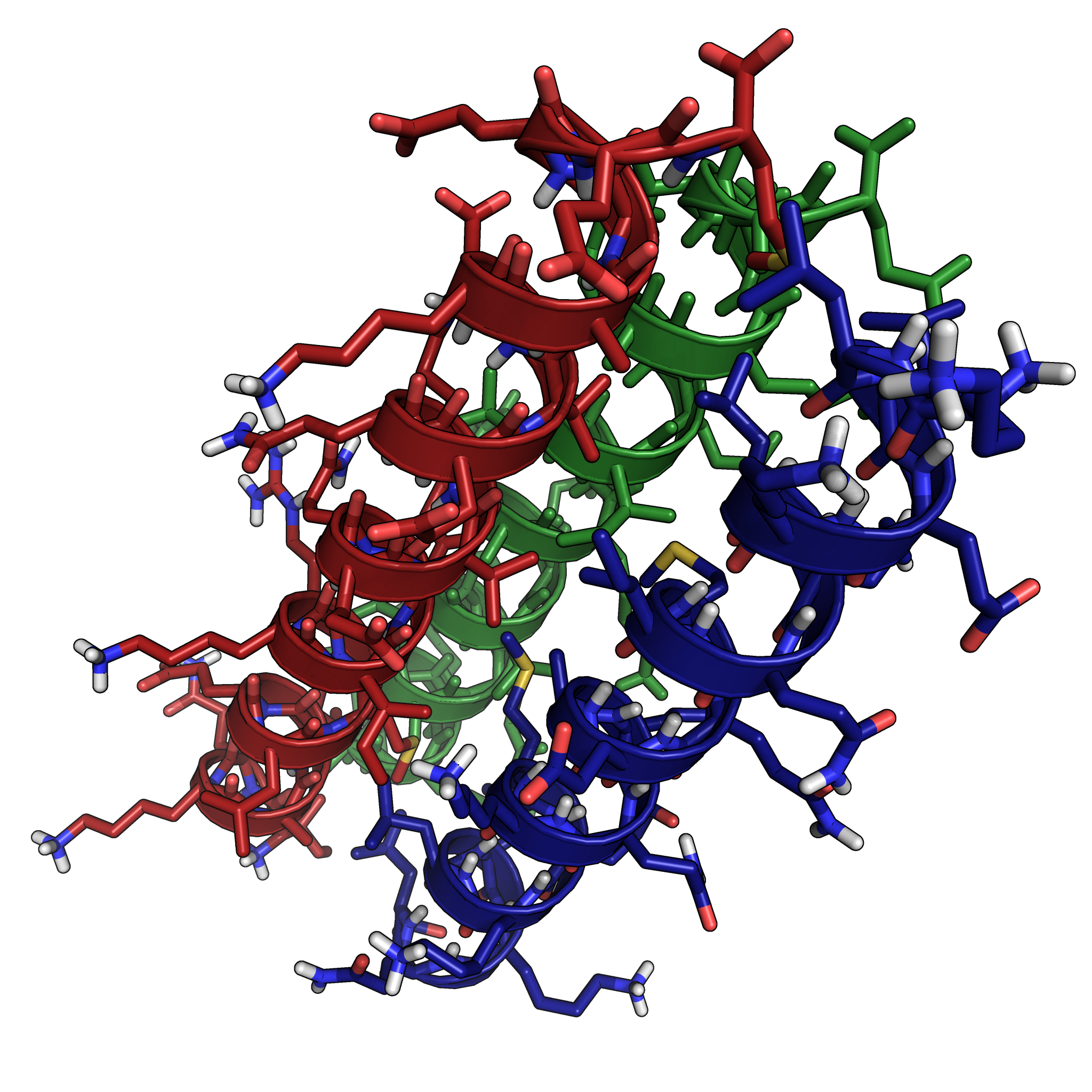}
    \includegraphics[width=5.5cm]{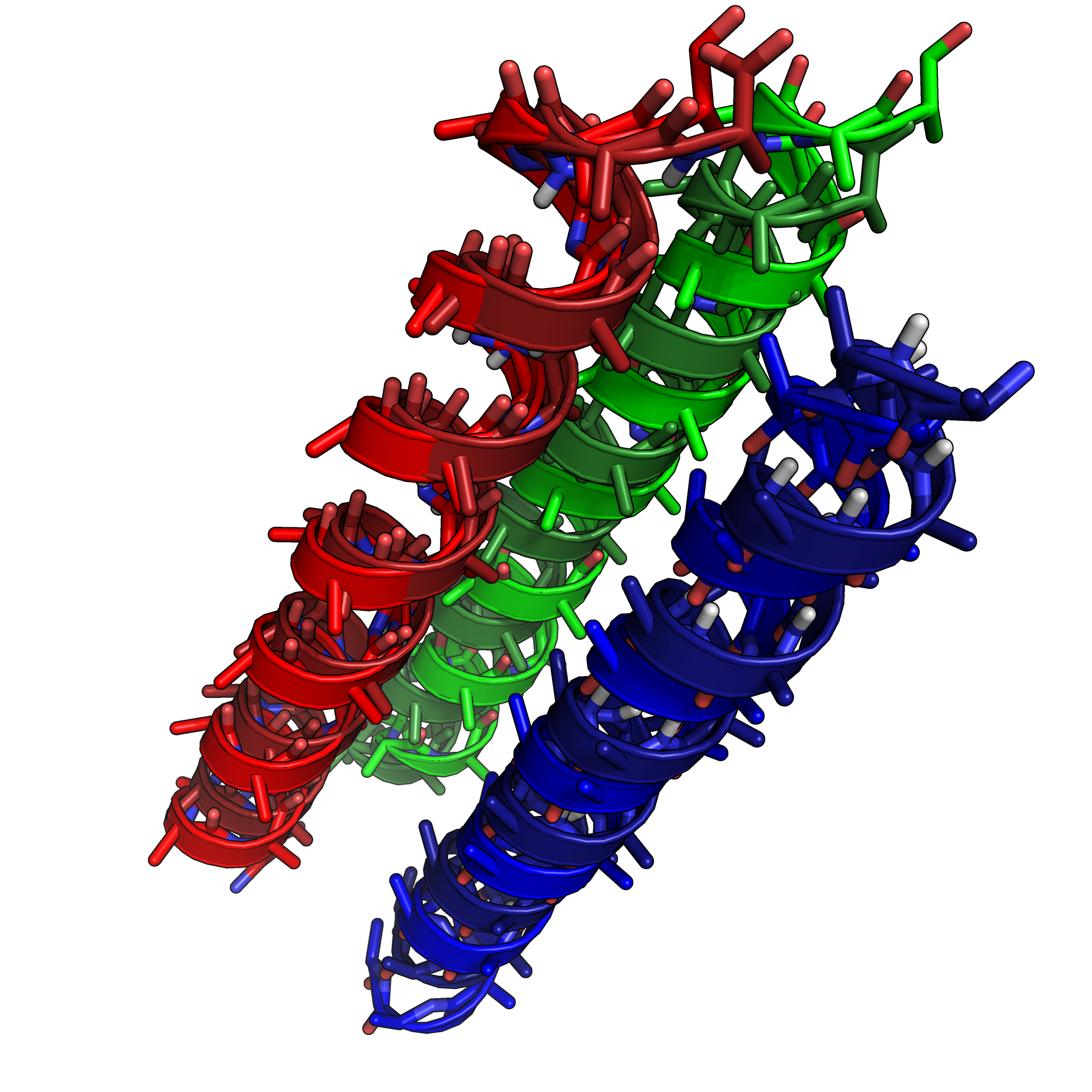}
   \caption{Comparison of actual structure (left panel) and predicted lowest-energy structure from the quantum annealer (middle panel, with main-chain overlay in right panel).  The three helices are shown in red, blue, and green respectively, with darker colours for the prediction.  The mainchain heavyatom RMSD between prediction and actual structure was 2.4 $\AA$, which is less than the 3.0 $\AA$ spacing of the sampled grid.}
   \label{fig:PredActualComparison}
\end{figure*}%[ht!]

As described in section \ref{Problem setup}, we constructed a simple three-body docking problem ($N=3$) to test our method by dividing an $\alpha$-helical protein into three helical fragments.  In order to fit the problem to the currently-available quantum annealing hardware, we fixed one of the helices and sampled 49 points ($T=49$) in the two-dimensional physical space for each of the other two helices. The space sampled was a plane perpendicular to the bundle axis. Given a fixed helix, the sampling problem at hand corresponds to finding the lowest energy configuration in a twelve-dimensional solution space, reduced to four dimensions by considering only translations in the plane of the other two helices. %covering a 2-dimensional subspace of the 6-dimensional relative translation and rotation space. 

 Of the $T^{N-1}=2,401$ possible solutions to this small three-body docking problem, exhaustive enumeration revealed that the lowest-energy solution had an RMSD of 2.4 $\AA$ from the native structure of the protein as determined by x-ray crystallography; that is, given that the translational grid spacing was 3.0 $\AA$, the global minimum for this problem given the energy function used does indeed correspond to the true minimum to the limits of sampling resolution.  We found that with greedy post-processing, the D-Wave quantum annealer was able to find this solution reliably with fewer than 100 samples when hyperparameters were optimally tuned (see section \ref{R&D:quality} for full details), providing a simple proof of principle that problems of this class can be solved by quantum annealing.  Note that although this problem was simplified by using sparse grid sampling on a subspace of the full twelve-dimensional solution space, no simplification was made to the representation of the molecule: energies were computed for the all-atom models of the helices, using the full Rosetta REF2015\_soft energy function.  Figure \ref{fig:PredActualComparison} shows the x-ray crystal structure of the portion of the protein modelled compared to the predicted docked configuration of the three $\alpha$-helices.

%\vmullig{Note that we can make a point that, given a 3 Angstrom voxel, the distance from the centre of the voxel to any corner is sqrt(3*1.5^2) = 2.6 A, so 2.4 A can be within the voxel.  A really nitpicky reviewer might ask whether the solution is ACTUALLY within the voxel, I guess, but I'm disinclined to worry about that for the preprint.}

\subsection{Penalty strength influence on intermixing of physical and non-physical states}\label{R&D:mixing}

%\mpand{In this subsection, we are going to show what's the affect of tuning one-hot penalty strength. What is the effect? Does it increase ruggedness of the spectrum or increase number of local minima? I am not sure what to emphasize here but maybe we can write this summary para after finishing the below analysis.}

\begin{figure*}
    \includegraphics[width=18cm]{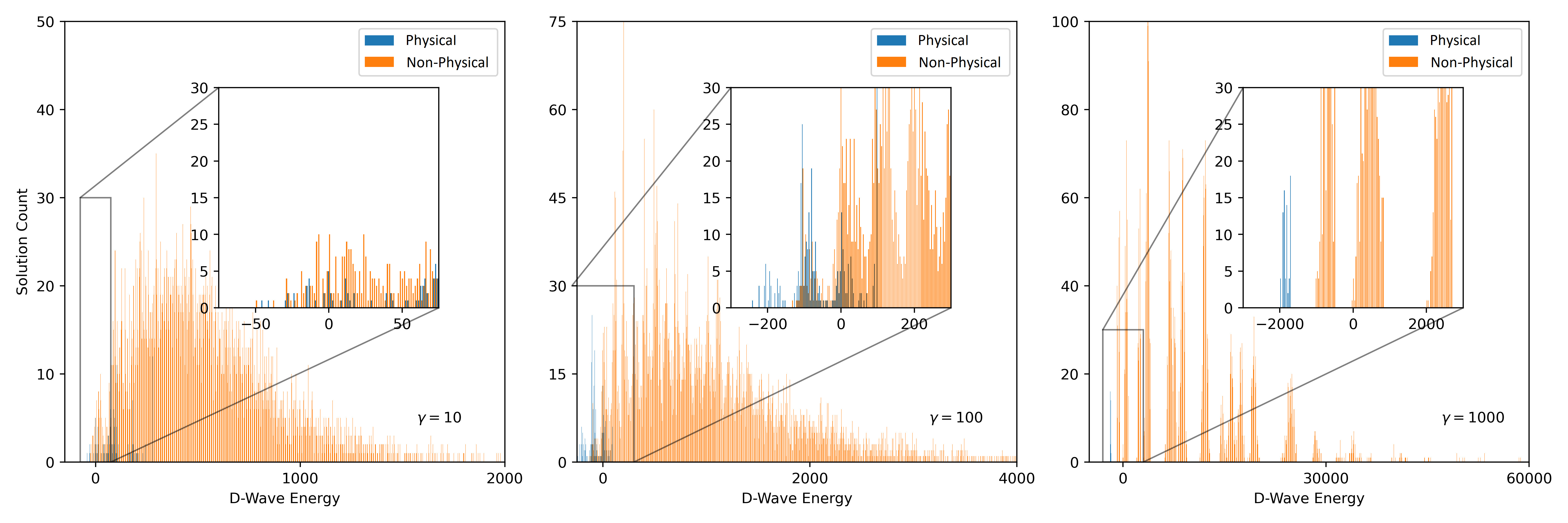}
    \caption{Sample solution energy distributions returned by D-Wave Advantage 4.1 of one-hot three helix bundle docking problem QUBO sampled $10^4$ times with annealing time $\tau = 1$ $\mu$s, for one-hot penalty strength $\gamma = 10$ (left panel), $\gamma = 100$ (central panel), and $\gamma = 1000$ (right panel). Physical and non-physical solutions are shown in blue and orange, respectively. When $\gamma = 10$ (small $\gamma$ regime), total mixing between physical and non-physical states occurs, so that the lowest and highest energy states are non-physical. When $\gamma = 100$ (moderate $\gamma$ regime), some mixing between physical and non-physical states occurs; note, however, that the lowest energy states are physical. When $\gamma = 1000$ (large $\gamma$ regime), physical and non-physical states are separated. In all cases, non-physical states are sampled with higher frequency than physical states, though sampling of low-energy physical states is most efficient in the moderate $\gamma$ regime.}
    \label{fig:QA_solution_histograms}
\end{figure*}

We consider Equation \ref{MBD_Hamiltonian} in three different regimes of the one-hot penalty strength, $\gamma$: (1) the \textit{large $\gamma$ regime}: when $\gamma$ is large enough that physical and non-physical states are separate, (2) the \textit{moderate $\gamma$ regime}: when $\gamma$ is such that there is mixing between physical and non-physical states, but some physical states are lower in energy than all non-physical states, and (3) the \textit{small $\gamma$ regime:} when $\gamma$ is such that there is mixing between physical and non-physical states, but some non-physical states are lower in energy than all physical states. In Figure \ref{fig:QA_solution_histograms}, the frequencies with which the D-Wave quantum annealer samples states as functions of state energy are shown for each of these cases.

In the large $\gamma$ regime, at the extreme where $\gamma\gg |E|$, with $|E|$ being the largest-magnitude energy term appearing in Equation \ref{MBD_Hamiltonian}, we get the following QUBO:

\begin{equation}
    H_{\text{MBD}} (\gamma \gg |E|) \approx \gamma \sum_{x\in X^{\prime}}\Big(-\sum_{i=1}^{T}b_i^x + 2 \sum_{i>j}b_i^xb_j^x\Big).
\label{MBD_Hamiltonian_large_gamma}
\end{equation}

% \begin{equation}
%     \gamma^*= {N\choose 2} E^+ - {(N-1)\choose 2} E^-
% \end{equation}

%In the $\gamma\rightarrow\infty$ limit, all $T^{N-1}$ physically-meaningful states have energy of $-\gamma (N-1)$ while non-physical states correspond to higher energy. This means that the energy landscape has global minima with $T^{N-1}$ degeneracy. As we reduce the value of $\gamma$, we uplift this degeneracy because of one- and two-body energy term.
%$2^{(N-1)T}-2^{(N-1) \log_2(T)}$
\noindent In this regime, all $T^{N-1}$ physical states have an approximate energy of $-\gamma (N-1)$, while all non-physical states correspond to higher energies, the lowest non-physical excited state having an approximate energy of $-\gamma(N-2)$ (see Appendix \ref{appendix:mixing} for details).  All local minima correspond to physical states, so that one can always move downhill by successive bit-flips from any non-physical state to a physical state in this regime. However, once in a physical state, it's impossible to move further downhill by local bit-flips.  (Note that moving in this way, one has no guarantee of finding a physical state \textit{of interest}, such as one of energy below a desired threshold.) As we reduce the value of $\gamma$, we enter the moderate $\gamma$ regime by uplifting the quasi-degeneracy of physical ground states, along with reducing the gap between physical states and the first excited non-physical state.

%The rest of spectrum is comprised of an exponential number of non-physical states with higher energies. 
%The ``hardness'' of the energy landscape is exemplified by the following thought experiment: starting with a random bit-string and following a greedy descent algorithm by carrying out local bit-flips, where in the energy landscape do we end up? Initially, we would be drawing a non-physical state with probability $(2^{(N-1)T}-T^{N-1})/2^{(N-1)T}$, and then would cross the energy barrier of $\gamma$ and land in one of the physical states but we won't be able to go downhill after that.
%Even though prior to greedy descent only a small proportion of the target diversity was sampled, a sufficient diversity of non-physical states was sampled in the sample limit of $10^4$ that the local minima comprising the target diversity were reached by greedy descent from these non-physical states.
%within the energy landscape
In the moderate $\gamma$ regime, only those physical states with energies lower than all non-physical states are guaranteed to be local minima. This has the effect of reducing the number of local minima in the landscape.  Due to the intermixing between physical and non-physical states, it is sometimes possible to move downhill from a physical state using local bit-flips by first moving to a non-physical state and then to a lower-energy physical state.  While this does not guarantee that low-energy physical states will be found, it increases the paths by which one could be found, and thereby increases the likelihood that these states will be found.

%For physical states greater in energy than the lower bound of the non-physical state distribution, it is sometimes possible to move to lower-energy physical states by local bit-flips by first moving to a lower-energy non-physical state.

%Therefore, in this regime, we expect to see greater convergence of solutions upon greedy post-processing to those physical solutions with energies lower than any non-physical states. This is consistent with our observations in Figure \ref{fig:QA_optimum_and_diversity}, which show that when $\gamma = 100$, which allows some energy mixing as shown in Figure \ref{fig:QA_solution_histograms}, faster convergence to full diversity is achieved versus when $\gamma = 1000$.

Upon entering the small $\gamma$ regime, certain non-physical states start to appear as local minima as they take on energies less than the energy of any physical state. Starting from either a physical state or a non-physical state higher in energy, it is often possible to move downhill by local bit-flips. However, one often gets stuck in a non-physical local minimum. Continuing to decrease $\gamma$, a threshold is reached whereupon all local minima are non-physical states.

Appendix \ref{appendix:mixing} describes in detail how to efficiently estimate the bounds of physical and non-physical state distributions as a function of $\gamma$ given only the energy scale and number of quantum registers of a one-hot QUBO. This allows one to estimate in advance the value of $\gamma$ required to separate fully between physical and non-physical states, for example, or the value required to expose a certain percentage of the physical state energy band, a useful first step in tuning the ruggedness of the energy landscape to suit the needs of a particular problem.

%Note that only a subset of those non-physical states with energies lower than all physical states are guaranteed to be local minima. 

%For the three helix bundle problem, $\gamma\leq10$ falls within this regime. Our observations show that, in fact, all physical states found prior to post-processing convert to non-physical states upon post-processing, reducing the target solution diversity to identically zero.

\subsection{Solution diversity}\label{R&D:diversity}

\begin{figure*}[ht!]
    \includegraphics[width=8cm]{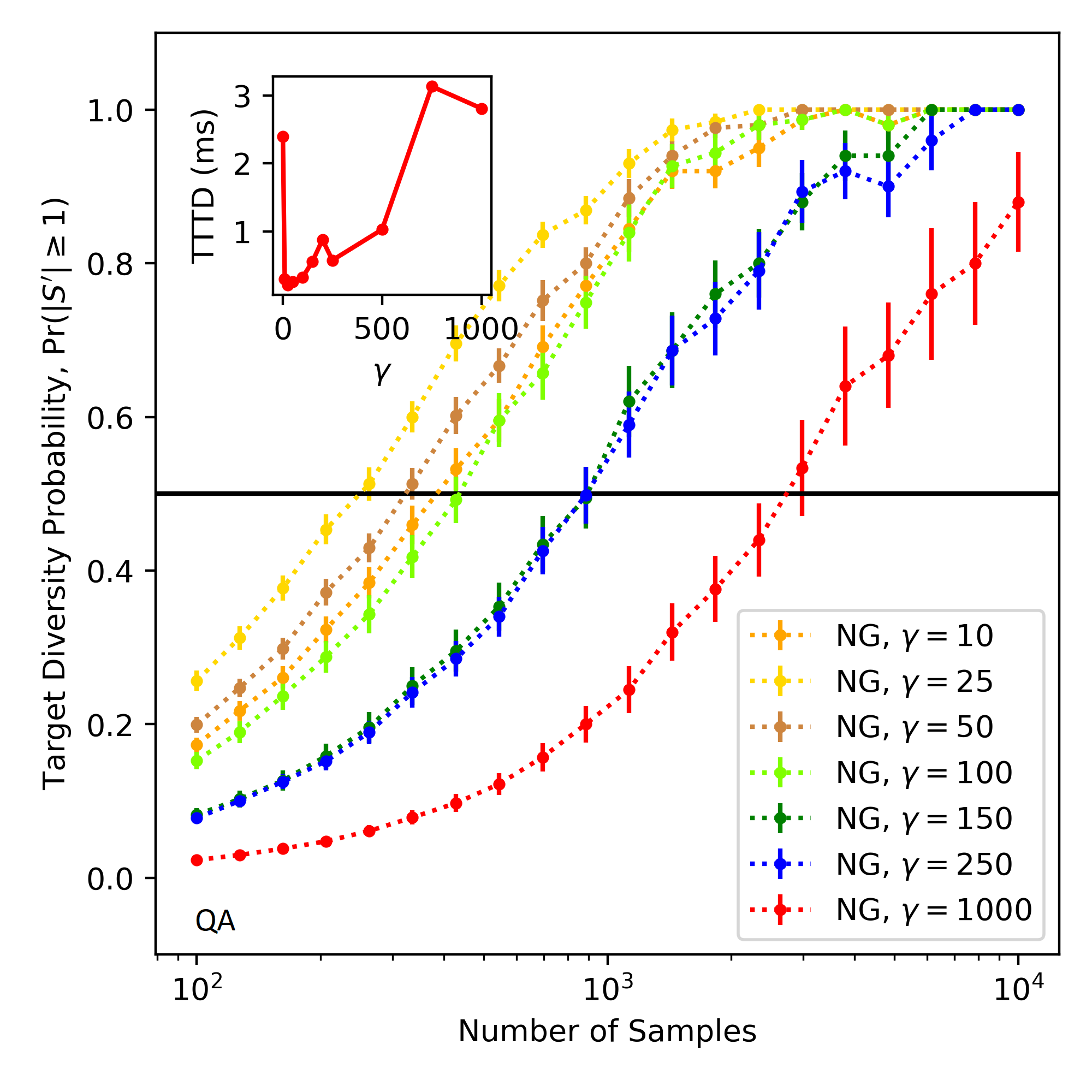}
    \includegraphics[width=8cm]{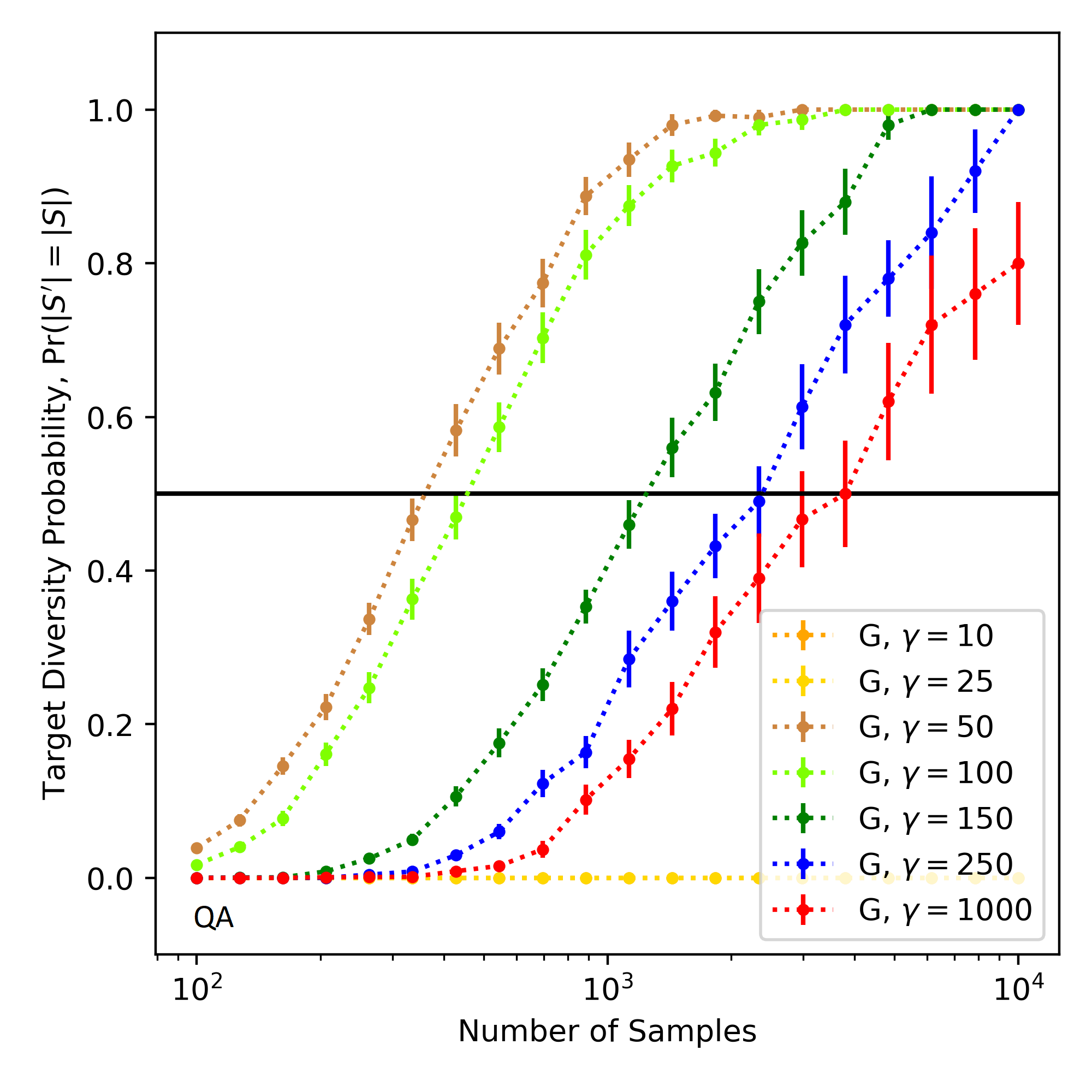}
    \includegraphics[width=8cm]{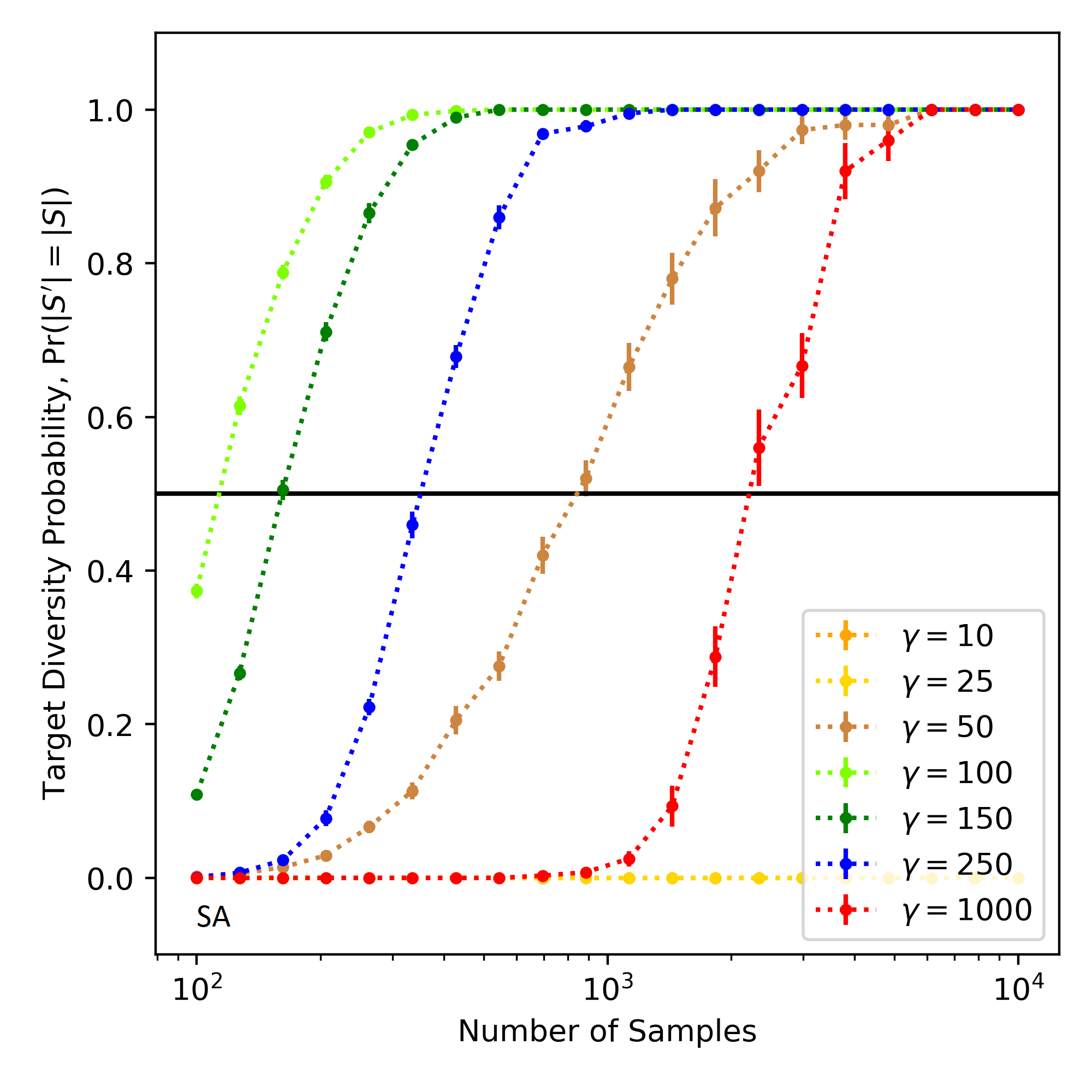}
    \includegraphics[width=8cm]{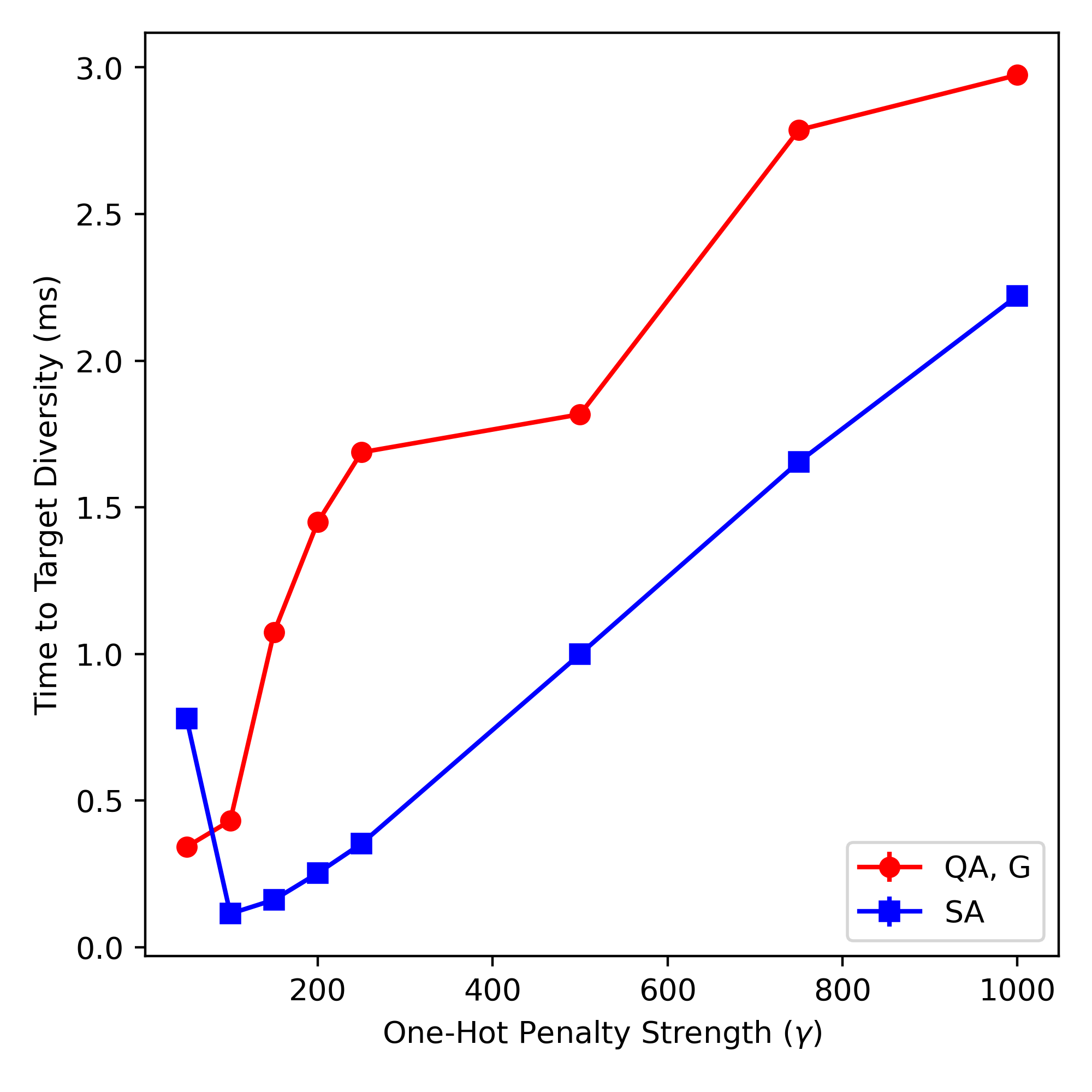}
\caption{(Upper left) Three helix bundle target diversity probability shown as a function of sample size without greedy post-processing (NG for non-greedy). On the y-axis, we have $\text{Pr}(|S^{\prime}|\geq 1)$, which is the probability of the D-Wave quantum annealer sampling a subset $S^{\prime}\subset S$ of at least one target state from the set of target states in the given number of samples. Inset: Time-to-target diversity (TTTD), based on 50\% probability of achieving target diversity, plotted as a function of $\gamma$.  TTTD is minimum for $\gamma=25$. (Upper right and lower left) Target diversity probability as a function of sample size for (upper right) quantum annealing with greedy post processing and (lower left) QUBO simulated annealing. On the y-axis, we have $\text{Pr}(|S^{\prime}| = |S|)$, which is the probability of the quantum annealer sampling the full target diversity.  Full target diversity is achieved with fewest samples for the two algorithms (greedy post-processing and QUBO simulated annealing) at $\gamma=50$ and $\gamma=100$, respectively. (Lower right) Quantum annealing with greedy post-processing and simulated annealing time-to-target diversity as a function of $\gamma$. Note that only the annealing time of 1 $\mu$s is counted toward TTTD in the case of quantum annealing. A single simulated annealing sample is estimated as 1 $\mu$s assuming each step of the 1000 steps constituting the anneal schedule to take 1 ns.}

%\mpand{I am making a list of edits that need to be done. Let's do it in one go on Monday rather than doing it multiple times. 1. How about we say 'a)',' b)' in the pics so that it's easier to refer to them in caption and text? 2. Should we add word "QA", "QA, greedy post-processed", "SA" in the three figures so that reader can understand the difference between the two plots? }}

\label{fig:QA_diversity}
\end{figure*}

In Figure \ref{fig:QA_diversity}, we present the diversity of low-energy physical states sampled by the D-Wave quantum annealer and QUBO simulated annealing sampler over the three regimes of one-hot penalty strength $\gamma$, which was discussed before. These regimes correspond to various degrees of mixing between physical and non-physical states.

Focusing first on the solutions returned by the D-Wave quantum annealer prior to greedy post-processing (non-greedy), we investigate target diversity through the probability of the sampler returning at least one state from the target set within a given number of samples. We see that in starting from small $\gamma$ ($< 10$), as we increase $\gamma$ slightly to $10\leq\gamma\leq50$, we decrease the number of samples required to achieve full diversity. Increasing $\gamma$ beyond $50$ to $100$ serves to worsen the time required to achieve full diversity. We see a similar trend in average normalized diversity, \textit{i.e.} the average number of negative energy states divided by the number of target states (Figure \ref{fig:diversity_vs_samples}, left panel, and inset Figure \ref{fig:diversity_vs_gamma}, left panel).

Upon greedy post-processing, some to all target states are brought to either other lower-energy target states, or to non-physical states, for those values of $\gamma$ tested less than 50 (1, 10, 25). As a result, we see that the full diversity is unachievable for these values of $\gamma$. Figure \ref{fig:diversity_vs_gamma} (left panel) best demonstrates this effect, showing that for $\gamma\leq10$ in particular, no target states are found with greedy post-processing, while for $\gamma\geq50$, all target states are found in the large sample limit. This indicates that $\gamma\leq10$ lies in the small $\gamma$ regime, while $\gamma\geq25$ lies in the moderate $\gamma$ regime, where $\gamma\geq50$ is required to fully expose the physical states corresponding to the target set $S$. We see that at this value of $\gamma=50$, $S$ is fully sampled more rapidly compared to greater values of $\gamma$ as seen from the trend of normalized diversity as a function of $n$ samples (Figure \ref{fig:diversity_vs_gamma}, left panel). 

Interestingly, we note that between non-greedy and greedy post-processing of the quantum annealer solutions, different values of $\gamma$ demonstrate best performance as judged by the minimal number of samples required to achieve the target diversity with $50\%$ confidence. In the non-greedy case, $\gamma = 10$ demonstrates fastest convergence to the target set of states  (Figure  \ref{fig:QA_diversity}, upper-left panel), while in the greedy case, $\gamma = 50$ performs best (Figure  \ref{fig:QA_diversity}, upper-right panel). The latter we suggest should be true due to the following: when $\gamma = 50$, as discussed, the physical part of the spectrum of states corresponding to the target set becomes fully exposed, so that the energy landscape then essentially consists in 13 local minima, each corresponding to a target state. All other sampled states, which are preferentially non-physical states (see Figure  \ref{fig:QA_solution_histograms}), are brought by greedy descent with overwhelming likelihood to one of the 13 target states by an average of 5 iterations of the greedy descent optimizer. As $\gamma$ increases, more local minima are exposed, so that a given bit-string will be optimized by greedy descent with less likelihood to one of the target states. For $\gamma < 50$, there are fewer physical minima than the target states, which makes it impossible to reach the full target set.

For QUBO simulated annealing, as with greedy quantum annealing, all samples for $\gamma\leq25$ represent non-physical states, or otherwise a subset $S^{\prime}\subset S$, so that full target diversity is never achieved within this regime.  QUBO simulated annealing fully samples $S$ at $\gamma = 50$; however, the required minimal number of samples to achieve full target diversity is higher than in the range of $100\leq\gamma\leq250$ (Figure  \ref{fig:QA_diversity}, lower-left panel) . In contrast, greedy quantum annealing produces the shortest time (minimal number of samples) to sample S at $\gamma = 50$.  We suspect that the difference between the two methods arises because when $\gamma = 50$, simulated annealing converges to the exact ground state more frequently when compared to D-Wave sampling due to hardware noise on the latter. To sample better the target diversity with simulated annealing, one needs to increase the height of the barriers between local minima by increasing $\gamma$, so that with a higher probability the simulated annealer gets stuck in the higher-energy states that constitute the target diversity. However, increasing $\gamma$ beyond the range of $100\leq\gamma\leq250$ decreases the probability of achieving the target set, since the probability of the simulated annealer being stuck in non-target minima increases.

In comparing greedy quantum annealing and QUBO simulated annealing time-to-target diversity (Figure \ref{fig:QA_diversity}, bottom-right panel), under the clock-time assumptions made for each, QUBO simulated annealing appears to outperform greedy quantum annealing by a constant factor of approximately 1.5 over most values of $\gamma$ tested, for the three helix bundle problem. However, at this stage, no definitive conclusions might be drawn concerning the general performance of these two methods, or their scaling. The latter requires that they be assessed relative to a continuum of problem scales, which within the context of molecular docking entails varying the number of docking bodies ($N$). It should be noted as well that, although $\gamma = 400$ approximates the common recommendation that one should choose a one-hot penalty strength about four times the maximum energy of the problem QUBO, we see an approximately 4-fold time savings in the case of greedy quantum annealing, and a 4-fold time savings in QUBO simulated annealing in comparing time-to-target diversity between optimized $\gamma$ and $\gamma = 400$. %\mpand{Should it not be $\gamma=400$? $4 times 100=400$ }

These investigations suggest a different ``sweet spot'' for $\gamma$ under each method of sampling with respect to the desired target diversity, all of which, nevertheless, require intermixing between physical and non-physical states. In the case of greedy quantum annealing and QUBO simulated annealing, due to the steepest-descent step that forms the end stage to each (effected through the low-temperature stage in simulated annealing), it is the ruggedness of the landscape and how widely this landscape is sampled that combine to determine the number and distribution of local minima achieved. In general, larger $\gamma$ equates to more physical  local minima and larger energy barriers between them. In the case of non-greedy quantum annealing, non-minima are sampled in addition to local minima due to hardware noise, so that the optimal choice of $\gamma$ is more a reflection of which energy band is preferentially sampled by the D-Wave solver. In our example, this energy band required full mixing between physical and non-physical states. It remains to be studied over a range of problems how to specifically select an optimal value of $\gamma$ for a given solver, but this work suggests that for each case energy mixing is required.

%\tz{Possible note on Vikram's idea that we track the states that simulated annealing searches through as it explores? Would this help convergence to diversity? Probably not in QUBO simulated annealing...}

\subsection{Solution quality}\label{R&D:quality}

\begin{figure*}[ht!]
    \includegraphics[width=8cm]{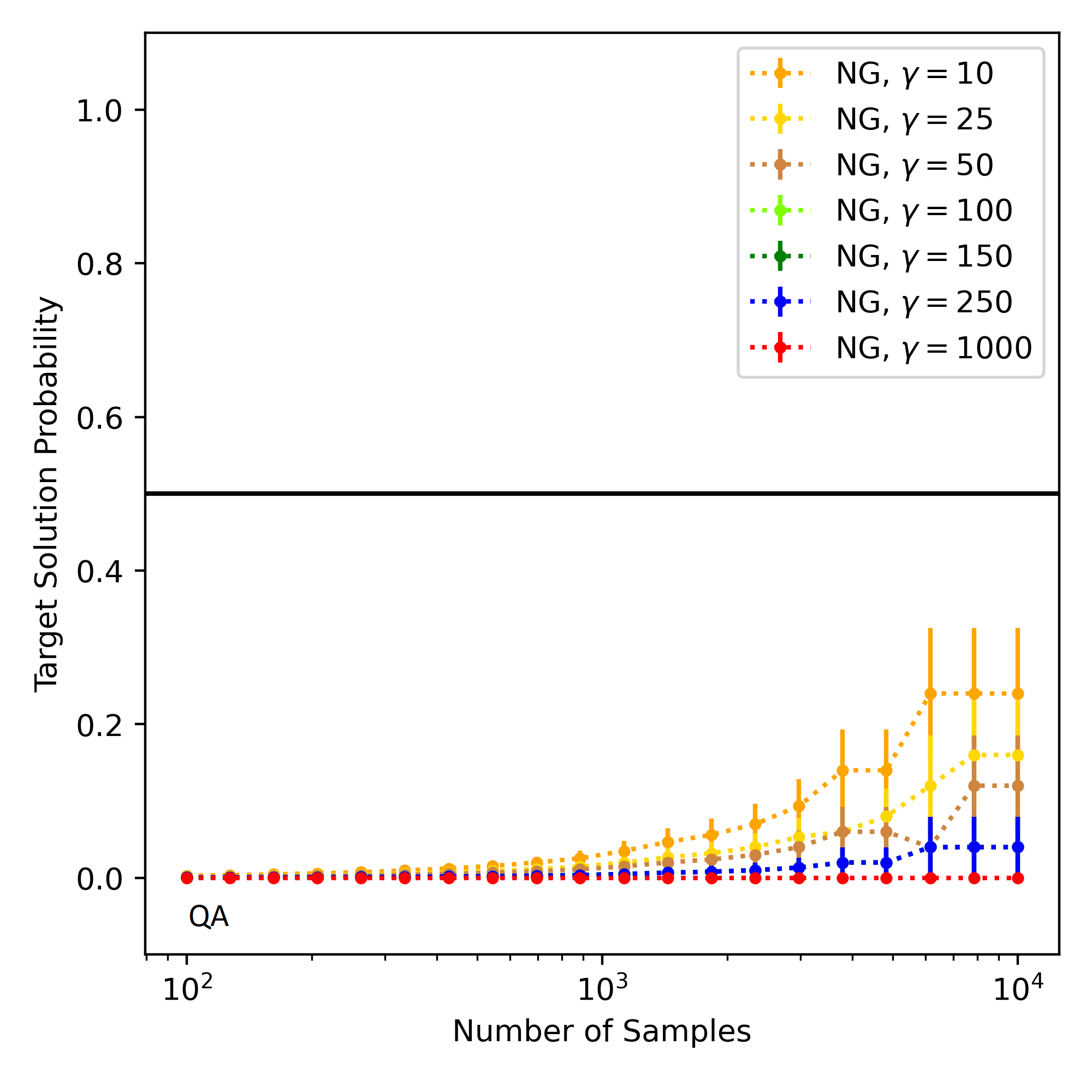}
    \includegraphics[width=8cm]{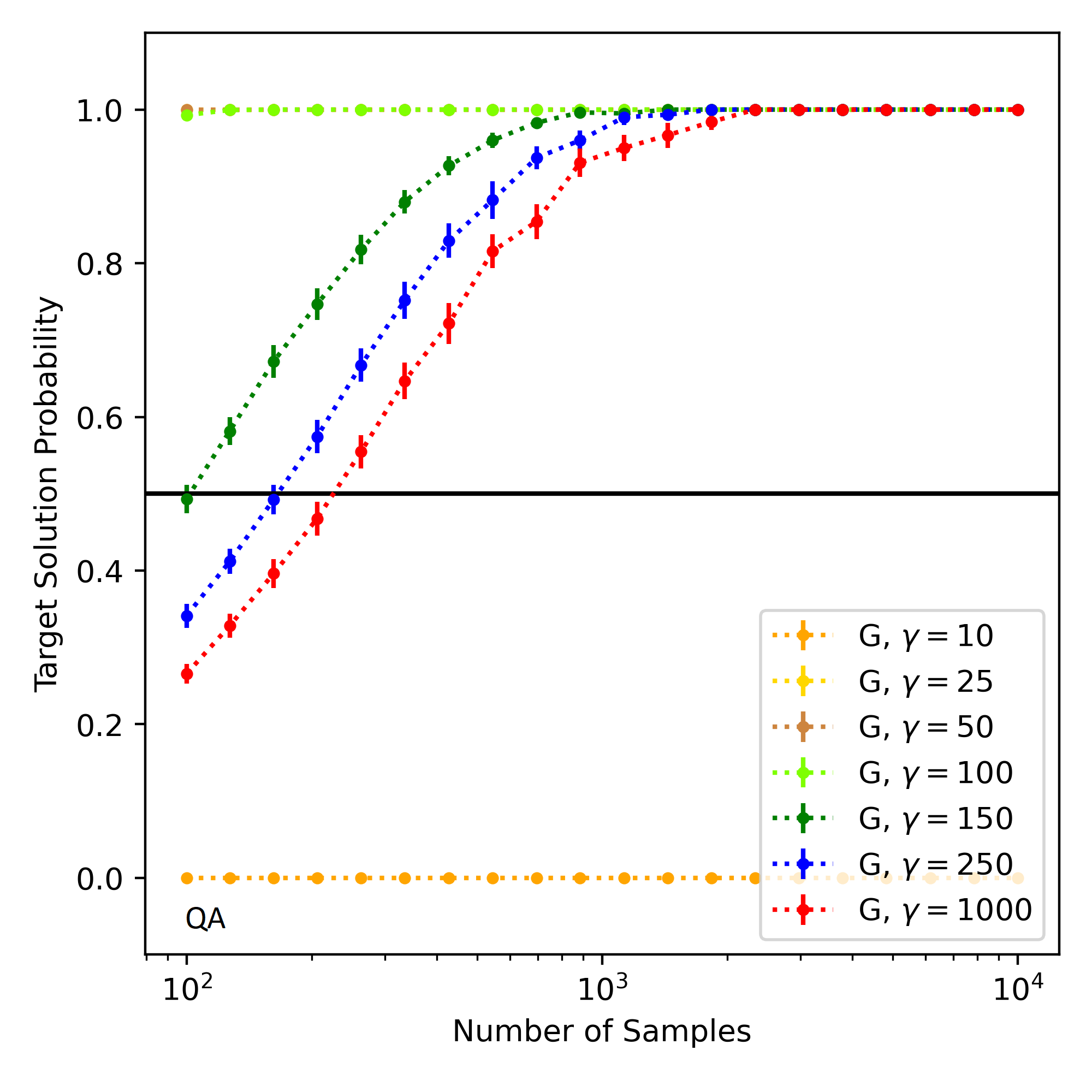}
    \includegraphics[width=8cm]{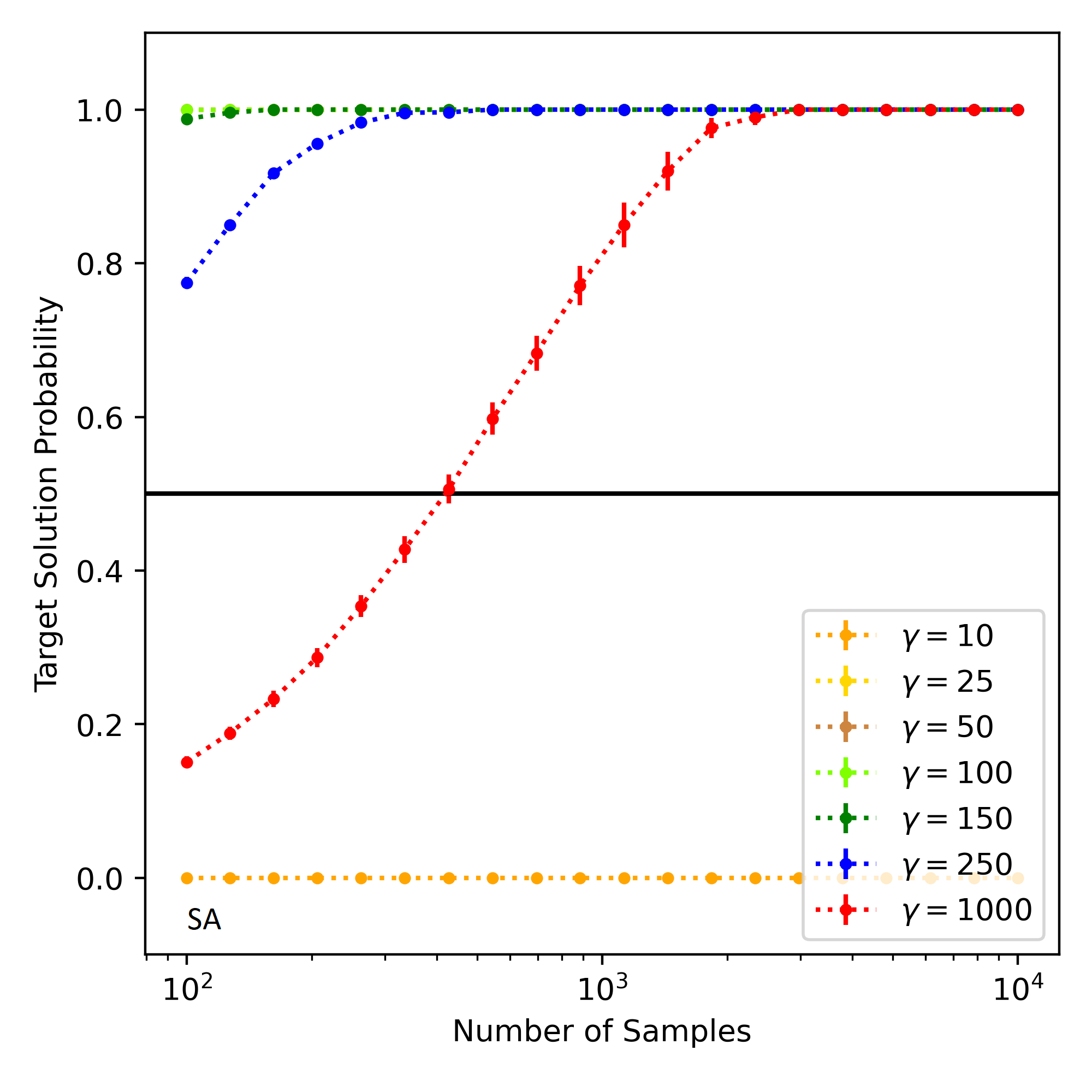}
    \includegraphics[width=8cm]{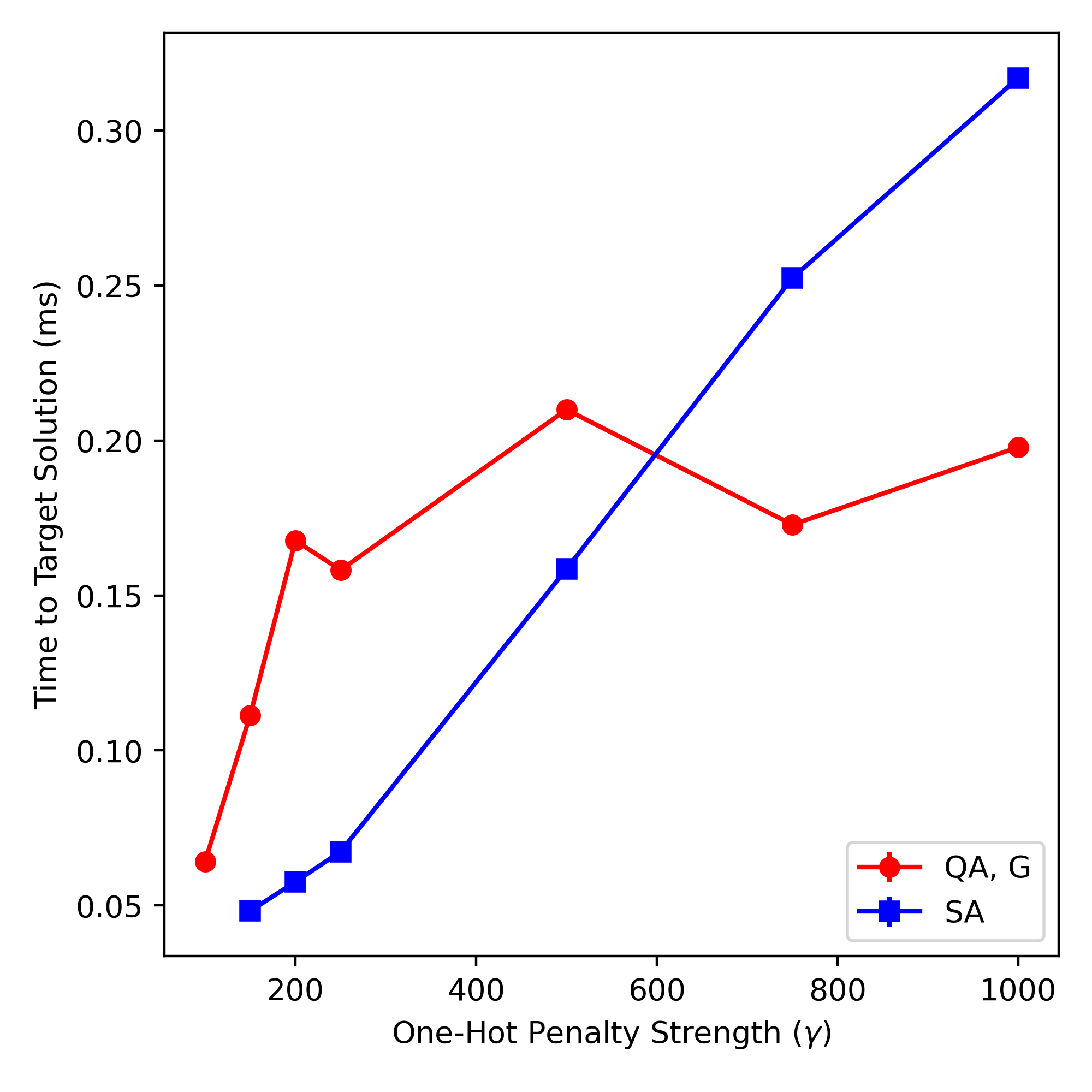}
    \caption{Three helix bundle ground state target solution probability as a function of number of samples for: (upper left) quantum annealing without greedy post-processing (NG for non-greedy), (upper right) quantum annealing with greedy post-processing, and (lower left) QUBO simulated annealing. Panels contain solver labels in their lower-left corners. (Lower right) Quantum annealing with greedy post-processing and QUBO simulated annealing time to target solution as a function of $\gamma$. Note that only the annealing time of 1 $\mu$s is counted toward TTTD in the case of quantum annealing. A single simulated annealing sample is estimated as 1 $\mu$s assuming each step of the 1000 steps constituting the anneal schedule to take 1 ns. %\mpand{Should we add labels such as SA and QA in the figures?}
    }
    \label{fig:QA_quality}
\end{figure*}

In Figure \ref{fig:QA_quality}, we present the relative performance of the D-Wave quantum annealer and QUBO simulated annealing sampler as measured by their ability to sample the ground state for our problem over a range of one-hot penalty strengths.  As observed for solution diversity, non-greedy quantum annealing (Figure \ref{fig:QA_quality}), upper-left panel) demonstrates best performance with respect to the known ground state when $\gamma = 10$, which occupies the small $\gamma$ regime of strong intermixing between physical and non-physical states. Increasing $\gamma$ reduces the probability of achieving the ground state in monotonic fashion until $\gamma = 1000$, where the ground state solution is never sampled. We expect that for $\gamma>1000$ we should continue to see failure to sample from the ground state due to limited dynamic range causing increased quasi-degeneracy of physical states and an increase in the effect of hardware noise. Note that over all $\gamma$, unit probability of sampling the ground state over the sample range tested is not achieved; at best, when $\gamma = 10$, a probability of 0.23 of sampling the ground state over $10^4$ samples is achieved. 

In the case of quantum annealing with greedy post-processing (Figure \ref{fig:QA_quality}, upper-right panel), for $\gamma>10$, the ground state solution is reached with unit probability in the large sample limit, with smaller values of $\gamma$ demonstrating faster convergence. Note that while best performance over the set of 13 diverse target solutions required $\gamma = 50$, here, the smaller value of $\gamma = 25$ is optimal. This corresponds to less exposure of the physical spectrum. In other words, a smaller number of local physical minima are present, which increases the probability that a steepest descent optimizer should bring a sampled bit-string to the ground state. Indeed, for $\gamma\leq100$, near-unit probability of sampling from the ground state is achieved in only 100 samples. For $\gamma>100$, a substantially greater number of samples is required to guarantee unit probability of sampling the ground state. As discussed in section \ref{R&D:diversity}, this dependence on $\gamma$ is a result of the number of local minima increasing with $\gamma$, so that a sampled bit-string is less likely to converge to the ground state with greedy descent.  It is expected that as problems grow larger any more complex, with larger $T$ (number of gridpoints per body) and $N$ (number of bodies), the probability of a random bitstring converging to the global optimum with greedy descent would drop considerably at any value of $\gamma$, so that while the tuning of $\gamma$ is important, greedy descent alone cannot be relied upon.

The probability of finding the optimal solution as a function of sample size with QUBO simulated annealing is shown in the lower-left panel of Figure \ref{fig:QA_quality}. The performance of QUBO simulated annealing displays precisely the same monotonic dependence on $\gamma$ as does greedy quantum annealing, though faster convergence in terms of number of samples required is observed for all $\gamma < 750$ (Figure \ref{fig:QA_quality}, bottom-right panel). For $\gamma\geq750$, the greater number of samples required for QUBO simulated annealing suggests that quantum annealing samples preferentially from the low-energy well containing the ground state as its minimum in the large $\gamma$ regime versus QUBO simulated annealing. We remark that this plateauing for greedy quantum annealing constitutes an interesting departure from the roughly linear scaling in $\gamma$ seen for: (i) QUBO simulated annealing on the time-to-solution metric, and (ii) greedy quantum annealing and QUBO simulated annealing on the time-to-target diversity metric.

%We remark that this plateauing for greedy quantum annealing time-to-target solution constitutes an interesting departure from the roughly linear scaling in $\gamma$ seen  both greedy quantum annealing and QUBO simulated annealing time-to-target diversity, and QUBO simulated annealing time-to-target solution.

Again, as in the case of solution diversity, we observe that the one-hot formulation of our multibody molecular docking problem shows better performance when the one-hot penalty strength, $\gamma$, is tuned to promote mixing between physical and non-physical states. Specifically, we see a 3-fold time savings in the case of greedy quantum annealing, and a 3-fold time savings in QUBO simulated annealing in comparing between optimized $\gamma$ and $\gamma = 400$, a selection approximating the recommended practice that chooses a one-hot penalty strength about four times the maximum energy of the problem QUBO. Optimized $\gamma$, as with solution diversity, varies with solver. In the case of non-greedy quantum annealing, we observe that the same choice of $\gamma$ optimizes against both solution diversity and quality for the three helix bundle problem, while different $\gamma$ are optimal for greedy quantum annealing and QUBO simulated annealing. It is likely that varying the target diversity may require different $\gamma$ for non-greedy quantum annealing solution diversity and quality.

\section{Conclusion}

%VKM CONTINUE HERE

In this work, we have shown that the multibody molecular docking problem, critical to the computational drug discovery pipeline, can be mapped to a form executable by current- and future-generation quantum annealing hardware using a one-hot encoding scheme. This mapping involves a classical pre-computation of pairwise interactions followed by sampling executed on a quantum annealer. Since the classical pre-computation scales only quadratically in the number of docking bodies, our approach delegates the harder part, sampling low-energy states in an exponentially large solution space, to the quantum annealer.  Our approach is notable for being able to leverage any pairwise-decomposable energy function, including the highly optimized Rosetta REF2015 energy function, without loss of fidelity. 

Testing our method against a realistic three-body problem, we observe that quantum annealing is able to sample diverse low-energy configurations reliably, especially when coupled with greedy-descent post-processing. These configurations include the ground state matching the experimentally-determined configuration to within the spacing of our sampling grid. Importantly, we observe that one-hot penalty strengths that allow intermixing of physical and non-physical states can increase the probability of sampling from these configurations by approximately 3-4 times when compared to recommended penalty strengths that do not allow intermixing. We find that this is because one-hot penalty strength effectively tunes the intermixing of physical and non-physical states, which in turn tunes the number of local minima. As assessed by the three body problem tested, we observe comparable performance between quantum annealing and QUBO simulated annealing.

%preferentially forcing higher-energy minima to non-minima
In this work, we have replaced the $NP$-hard step of the multibody docking problem with quantum sampling that can conceivably show better-than-classical performance scaling as quantum hardware grows larger and more robust.  Since we have done this without sacrificing accuracy in the polynomial-time energy pre-computation, we hypothesize that our approach could offer performance advantages for large multibody docking problems over classical approaches in the future.  Because multibody docking is a rate-limiting step when validating designed drugs binding to their targets \textit{in silico}, particularly when bound water molecules are modelled explicitly, this work has the potential to greatly accelerate drug discovery pipelines.

\bibliographystyle{IEEEtran}
\bibliography{IEEEabrv, references}

\newpage

\appendix

\renewcommand{\theequation}{A.\arabic{equation}}
\setcounter{equation}{0}
\renewcommand{\thefigure}{A.\arabic{figure}}
\setcounter{figure}{0}

\section{Estimating the inter-mixing effects of one-hot penalty strength}\label{appendix:mixing}

The general one-hot multibody docking Hamiltonian for a set $X = \{1, 2, ..., N\}$ of bodies is:

\begin{align}
    H &= \sum_{x\in X}\sum_{i=1}^{\alpha_x}\big(O(b_i^x)-\gamma\big)b_i^x\notag\\
    &+\sum_{x\geq y}\sum_{i>j}(T(b_i^x, b_j^y) + \delta)b_i^xb_j^y
    \label{mdoh_simple_again}
\end{align}

This Hamiltonian admits internal, rotation, and translational degrees of freedom, where $\alpha_x$ is the number of binary variables representing the possible body-configurations of body $x\in X$.

Ignoring internal and rotational degrees of freedom, this Hamiltonian simplifies to:

\begin{align}
    H &= \sum_{x\in X^{\prime}}\sum_{i=1}^{\alpha}\big(O(b_i^x)-\gamma\big)b_i^x \notag\\
    &+\sum_{x\geq y}\sum_{i>j}(T(b_i^x, b_j^y) + \delta)b_i^xb_j^y
    \label{mdoh_simple_reduced}
\end{align}

$X^{\prime} = \{1, 2, ..., N-1\}$ is the set of mobile bodies (where we have selected body $N$ from $X$ to be fixed), and $\alpha$ is the number of translational grid-points accessible to any of these mobile bodies. The $O(b_i^x)$ terms then represent the pairwise energies of mobile body $x$ relative to the fixed body, and the two body terms $T(b_i^x, b_j^y)$ the pairwise energies between mobile bodies $x$ and $y$. Recall that $\delta = 2\gamma$ if $x=y$ and is $0$ otherwise.

In this section, we derive upper and lower bounds to the energies of the distributions of physical and non-physical states, first in the large $\gamma$ limit, and then for general $\gamma$.

\subsection{Large $\gamma$}

Taking the largest absolute value of the one- and two-body energies of Equation \ref{mdoh_simple_reduced} to be $|E|$, when $\gamma\gg |E|$, the Hamiltonian simplifies to:

\begin{equation}
    H = -\sum_{x\in X^{\prime}}\sum_{i=1}^{\alpha}\gamma b_i^x+\sum_{x\geq y}\sum_{i>j}\delta b_i^xb_j^y
    \label{mdoh_simple_reduced_large_gamma}
\end{equation}

Considering first the minimum energy solution to this Hamiltonian, we can see that the double sum over $b_i^x b_j^y$, which adds only a positive contribution, can be avoided altogether if no greater than one variable from any register is selected (i.e., $1$). Note that by register $x$ we mean the set of variables $\{b_1^x, b_2^x, ..., b_{\alpha}^x\}$. Then, it is trivial to recognize that Equation \ref{mdoh_simple_reduced_large_gamma} is minimized when precisely $1$ variable is selected per register, such that the minimum energy solution is physical and with the energy $-\gamma(N-1)$. Incidentally, this is both the minimum and maximum energy of the distribution of physical states in the large gamma limit.

Considering now the maximum energy solution to this Hamiltonian, we recognize that this solution should correspond to the unary non-physical state, i.e., the state which is composed entirely of $1$s. This is because each time more than $1$ body is selected in a register, at least one contribution of $2\gamma$ is added to the cost of the solution, whereas only one contribution of $-\gamma$ is added. Then, evaluating Equation \ref{mdoh_simple_reduced_large_gamma} with this unary non-physical state yields an energy of $\gamma(N-1)(\alpha^2-2\alpha)$. 

We now identify the minimum energy non-physical solution as one which corresponds to a physical state, plus or minus one additional bit, located anywhere. Adding one additional bit contributes one $-\gamma$ term and one $2\gamma$ term to the cost (for a total contribution of $\gamma$), whereas subtracting one additional bit contributes one $\gamma$ term. It may be easily seen that to add or subtract 2 or more additional bits adds at least $2\gamma$. Given this, the minimum energy non-physical solution is then with an energy of $-\gamma(N-2)$.

In sum, when $\gamma\gg|E|$, the upper and lower bounds to the energies of non-physical and physical states are:

\begin{subequations}
    
    \begin{equation}
        \text{P}_{\text{UB}} = -\gamma(N-1) % \hfill ($P$, upper and lower bound)
    \end{equation}
    \begin{equation}
        \text{P}_{\text{LB}} = -\gamma(N-1) % \hfill ($P$, upper and lower bound)
    \end{equation}
    \begin{equation}
        \text{NP}_{\text{UB}} = \gamma(N-1)(\alpha^2-2\alpha) %\hfill ($NP$, upper bound)
    \end{equation}
   \begin{equation}
       \text{NP}_{\text{LB}} = -\gamma(N-2) % \hfill ($NP$, lower bound)
   \end{equation}
\end{subequations}

\noindent Where $\text{P}$ means physical, $\text{NP}$ means non-physical, $\text{UB}$ means upper bound, and $\text{LB}$ means lower bound.

\subsection{General $\gamma$}

We now seek the upper and lower energy bounds to the physical and non-physical state distributions when $\gamma$ is general. To carry out this analysis exactly would require that we fully solve the QUBO (given that the lower energy bound to the physical state distribution is the ground state), an impossibility for most actual problem instantiations, but we can estimate these bounds if we assume that: (1) the minimum energy pairwise interaction between any two bodies is $E^-$, and (2) the maximum energy pairwise interaction between any two bodies is $E^+$, where $E^+>E^-$. 

For real problems, we estimate $E^-$ as the minimum pairwise energy in the provided energy file, and $E^+$ as the maximum pairwise energy in the provided energy file. In practice, this maximum pairwise energy will usually be the supplied energy cap.

Now, we first consider a lower bound to the minimum energy physical state. In this case, we estimate that each pairwise interaction between bodies in this state is with the energy $E^-$. Evaluating Equation \ref{mdoh_simple_reduced} under this assumption then yields $(E^--\gamma)(N-1)+E^-{N-1\choose2}$, which simplifies to ${N\choose2}E^--\gamma(N-1)$.

Estimating an upper bound to the maximum energy physical state, we assume that each pairwise interaction between bodies in this state is $E^+$. Similarly to above, we evaluate Equation \ref{mdoh_simple_reduced} under this assumption and find an energy of $(E^+-\gamma)(N-1)+E^+{N-1\choose2}$, which simplifies to ${N\choose2}E^+-\gamma(N-1)$.

Considering now a lower bound to the minimum energy non-physical state, we identify this state similarly as before as one which corresponds to the minimum energy physical state, plus or minus one additional bit (whose pairwise energies with all other bodies is $E^-$). Under this assumption, evaluation of Equation \ref{mdoh_simple_reduced} gives an energy of ${(N-1)\choose 2}E^- -\gamma(N-2)$.

Finally, we estimate the upper bound to the maximum energy non-physical state by first recalling the unary state to be highest in energy, given previous arguments. This corresponds to each body besides the fixed body being placed at every location within the simulation space. This energy is then:

\begin{align}
    &\sum_{x\in X^{\prime}}\Big(\sum_{i=1}^{\alpha}\big(O(b_i^x)-\gamma\big)+\sum_{i>j}\big(T(b_i^x, b_j^x + 2\gamma\big)\Big) \notag\\ &+\sum_{x>y}\sum_{i>j}T(b_i^x, b_i^y)
\end{align}

Given that in practice we are with the energy files which contain each $O(b_i^x)$ and $T(b_i^x, b_j^y)$, we can evaluate this expression exactly. However, for the sake of argument, here we assume that we might estimate this evaluation by replacing each $O(b_i^x)$ and $T(b_i^x, b_j^y)$ with $\langle E \rangle = (E^++E^-)/2$, an approximation to the average pairwise energy between bodies. (In practice, this average likely skews in the direction of $E^+$ with energy capping.) Under this assumption, the maximum energy non-physical state is then with an energy of $(\langle E \rangle-\gamma)(N-1)\alpha+(\langle E \rangle+2\alpha)(N-1){\alpha\choose2}+\langle E \rangle {(N-1)\choose2}{\alpha\choose2}$.

When $\gamma$ is general, the upper and lower bounds to the energies of non-physical and physical states are:

\begin{subequations}
    \begin{equation}
        \text{P}_\text{UB} = {N\choose2}E^+-\gamma(N-1)
    \end{equation}
    \begin{equation}
        \text{P}_\text{LB} = {N\choose2}E^--\gamma(N-1)
    \end{equation}
    \begin{align}
        \text{NP}_\text{UB} &= (\langle E \rangle-\gamma)(N-1)\alpha \notag\\&+(\langle E \rangle+2\alpha)(N-1){\alpha\choose2}\notag\\&+\langle E \rangle {(N-1)\choose2}{\alpha\choose2}
    \end{align}
    \begin{equation}
        \text{NP}_\text{LB} = {(N-1)\choose 2}E^- -\gamma(N-2)
    \end{equation}
\end{subequations}

\begin{figure}
    \includegraphics[width=8cm]{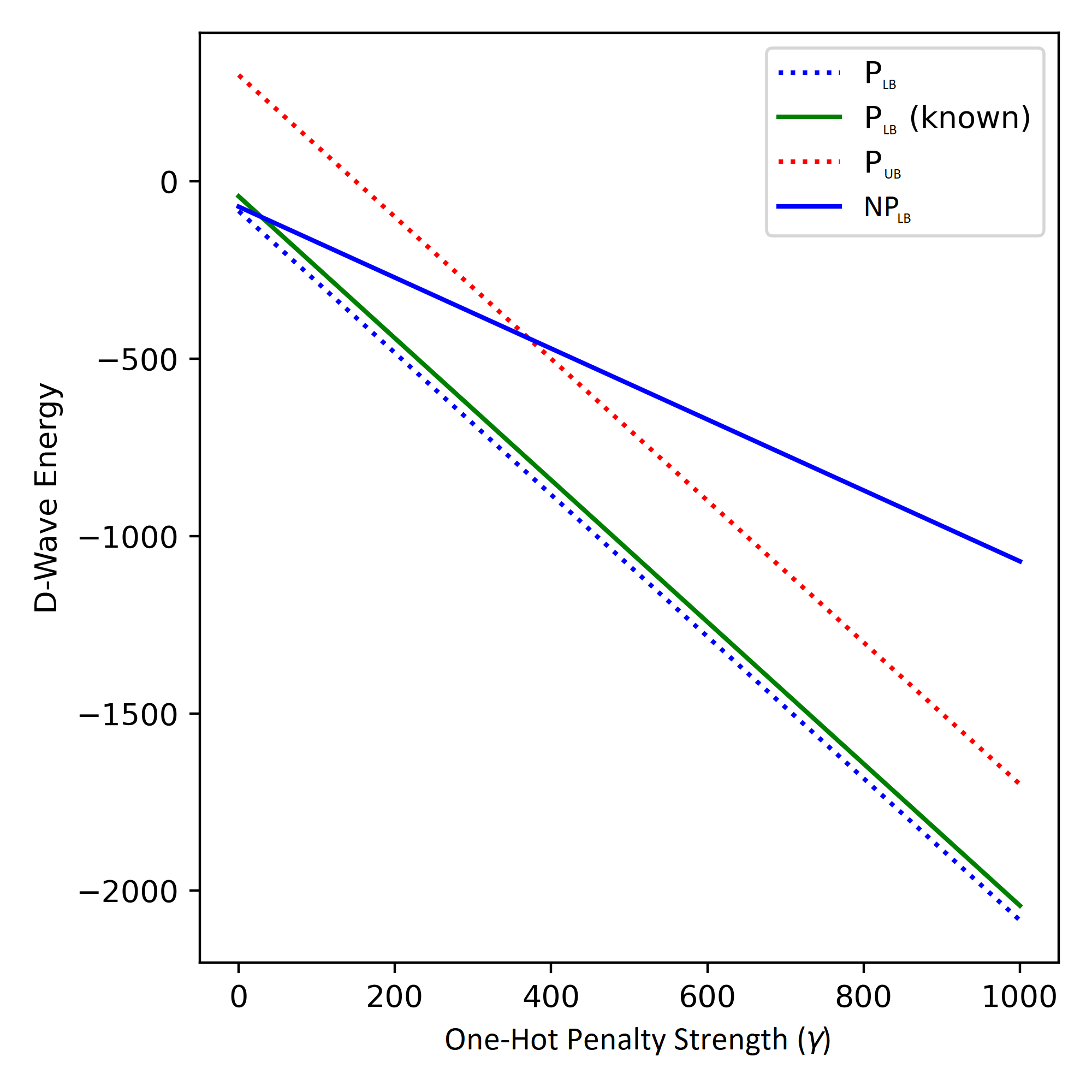}
    \caption{Predicted physical and non-physical state distribution energy bounds for three helix bundle (PDB ID: 4TQL) over a simulation-space confined to an $xy-$plane. The corresponding pairwise energy file was found to have a minimum pairwise interaction energy of $-38.83$ Rosetta Energy Units (REU), maximum pairwise interaction energy of $100.00$ REU, and a mean pairwise interaction energy of $22.26$ REU. Empirically, we have seen that the non-physical lower bound crosses the physical lower bound at $\gamma>0$, in contrast to prediction, but we do correctly predict that energy separation between physical and non-physical states occurs when $100\leq\gamma\leq1000$.}
    \label{fig__bounds_energy_vs_gamma}
\end{figure}

In Figure \ref{fig__bounds_energy_vs_gamma}, these relations are shown for fixed $N$, $E^+$, $E^-$, $\langle E \rangle$, and $\alpha$ against variable $\gamma$. The fixed values were selected based on the three helix bundle tested in this work (PDB ID: 4TQL).

We suggest this approach as a guide to determine where separation between physical and non-physical states occurs, and approximate degrees of energy mixing. One simplifying assumption we made took each minimum pairwise interaction energy between any two bodies to be identical (based on the absolute minimum of this set of energies), but it is possible to use the actual minimum pairwise energies between any two bodies. 

%\textit{Should we also find upper bounds to the lowest physical and non-physical energies? This estimate could be as follows: take the minimum negative energy between any two bodies, and identify this as the minimum physical state energy. This assumption has the interpretation that just two bodies are docked, while the others are undocked (far away, and with $0$ REU pairwise interaction energies). Otherwise, we can just assume that all bodies are undocked, in which case the REU energy of the configuration is $0$.}

\newpage
\onecolumngrid
\section{Solution diversity}

\begin{figure*}[ht!]
    \includegraphics[width=7cm]{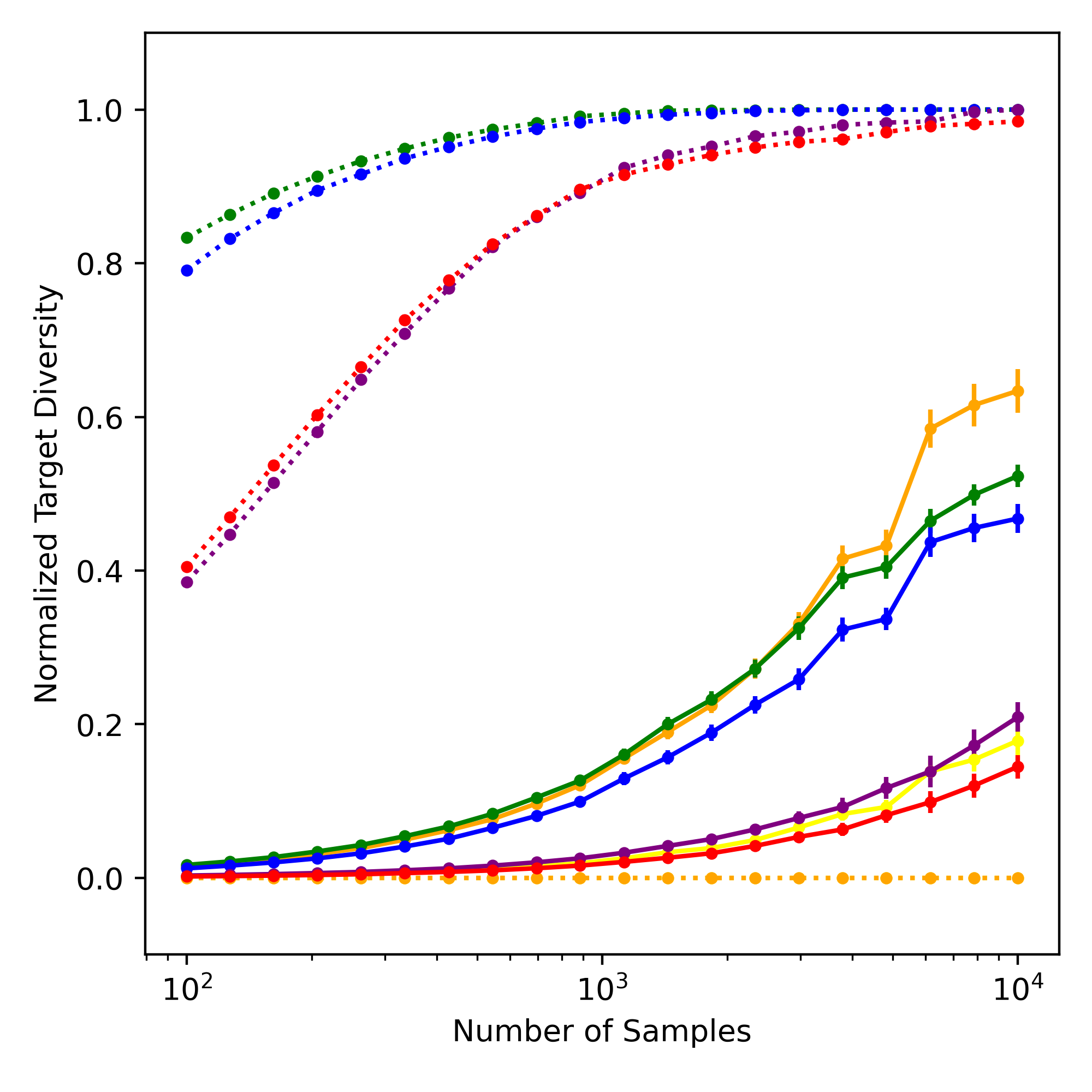}
    \includegraphics[width=7cm]{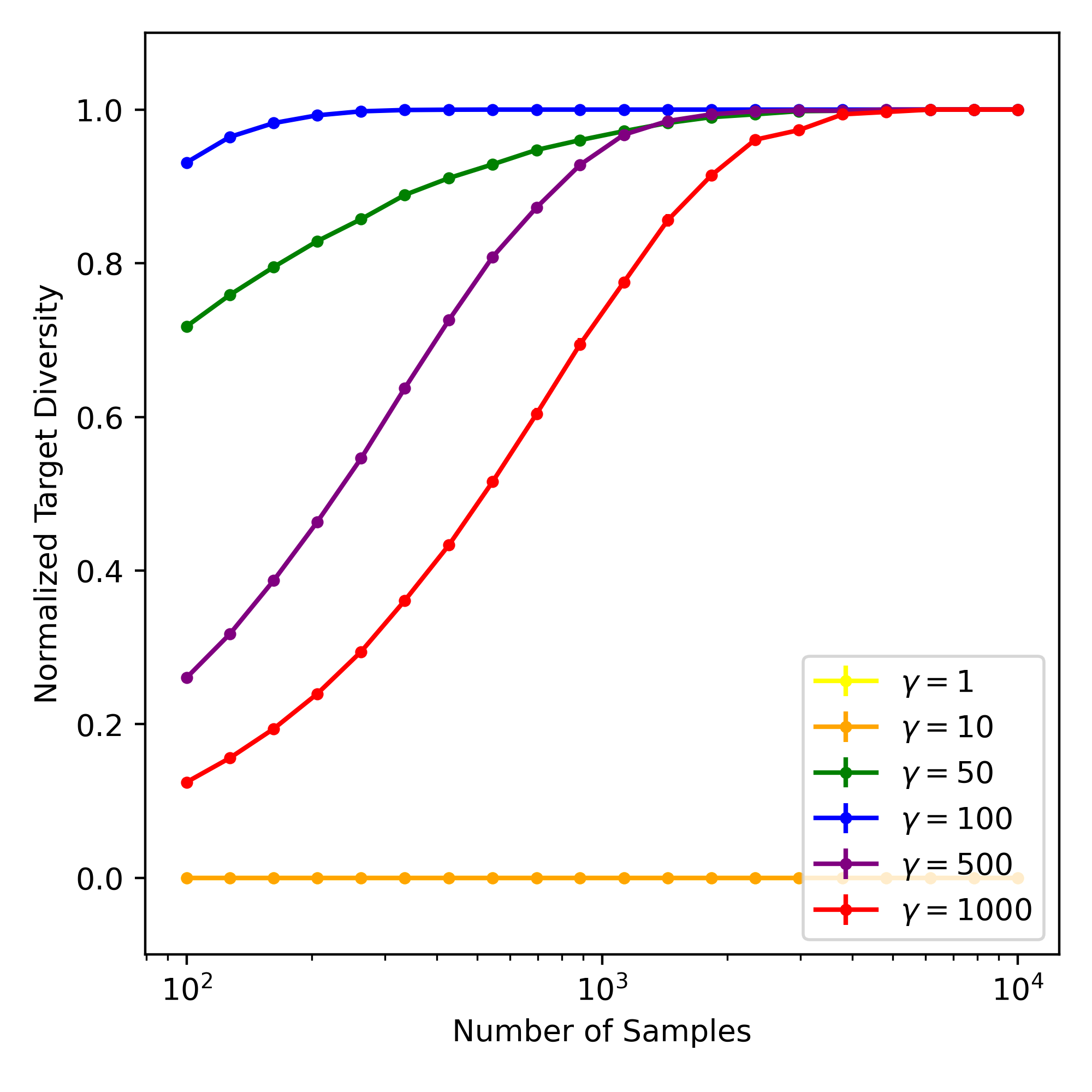}
    \caption{Average normalized target diversity (left panel) returned by D-Wave Advantage 4.1 of one-hot three helix bundle QUBO as a function of number of samples. Each point registers the mean of $25\cdot10^4/n$ replicates, where $n$ is the number of samples. Solid curves correspond to diversity prior to steepest descent post-processing (NG for non-greedy); dotted curves correspond to diversity with steepest descent post-processing (G for greedy). Non-greedy curves show better diversity for $1\leq\gamma\leq100$, while greedy curves show better diversity for $50\leq\gamma\leq100$. Greedy post-processing when $\gamma<50$ takes all solutions to non-physical states (zero diversity). The right panel shows average normalized diversity returned by the QUBO simulated annealing solver. Notice that for $\gamma = 100$ diversity is most quickly reached, and for $\gamma < 50$, all solutions are non-physical.}
    \label{fig:diversity_vs_samples}
\end{figure*}

\begin{figure*}[ht!]
\includegraphics[width=7cm]{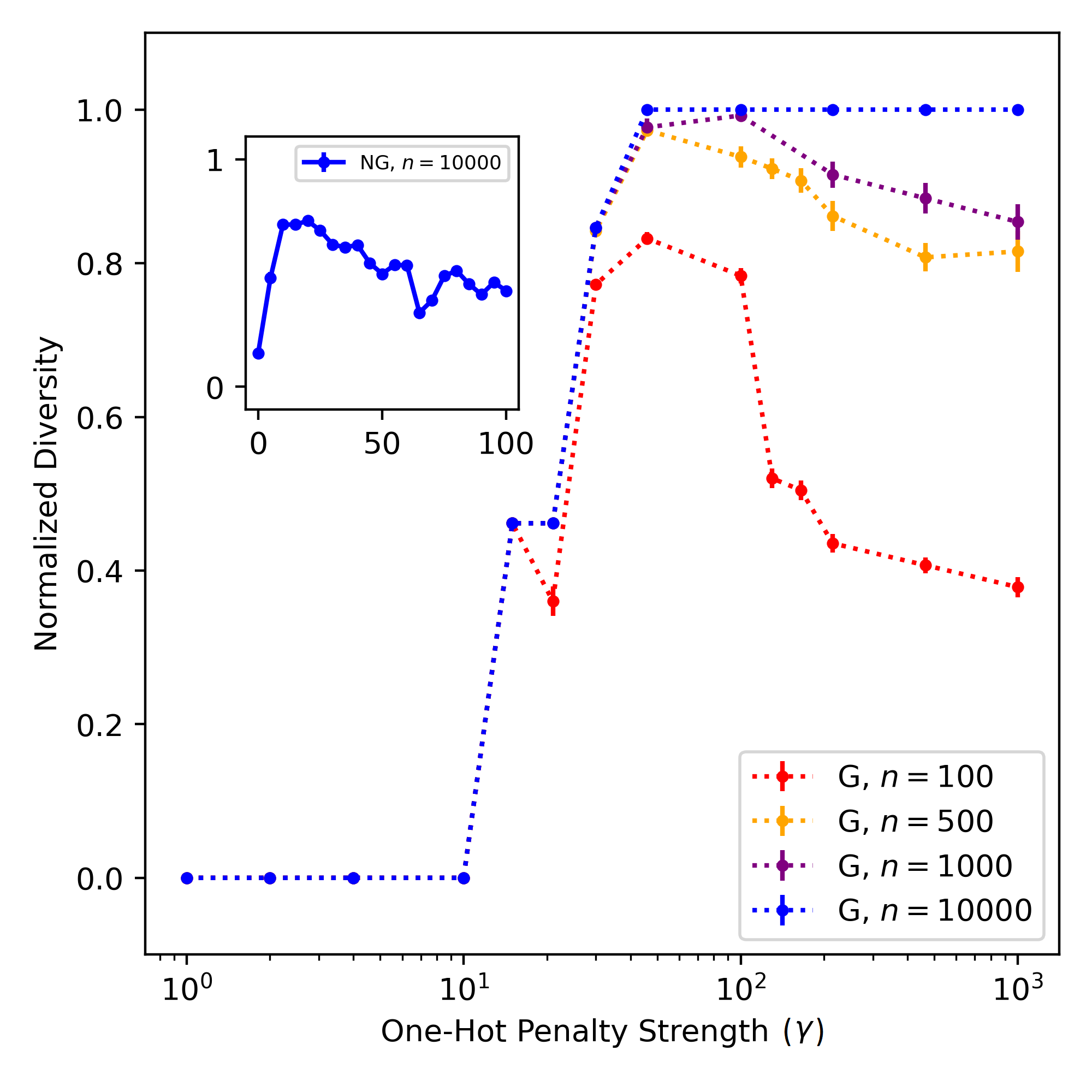}
    \includegraphics[width=7cm]{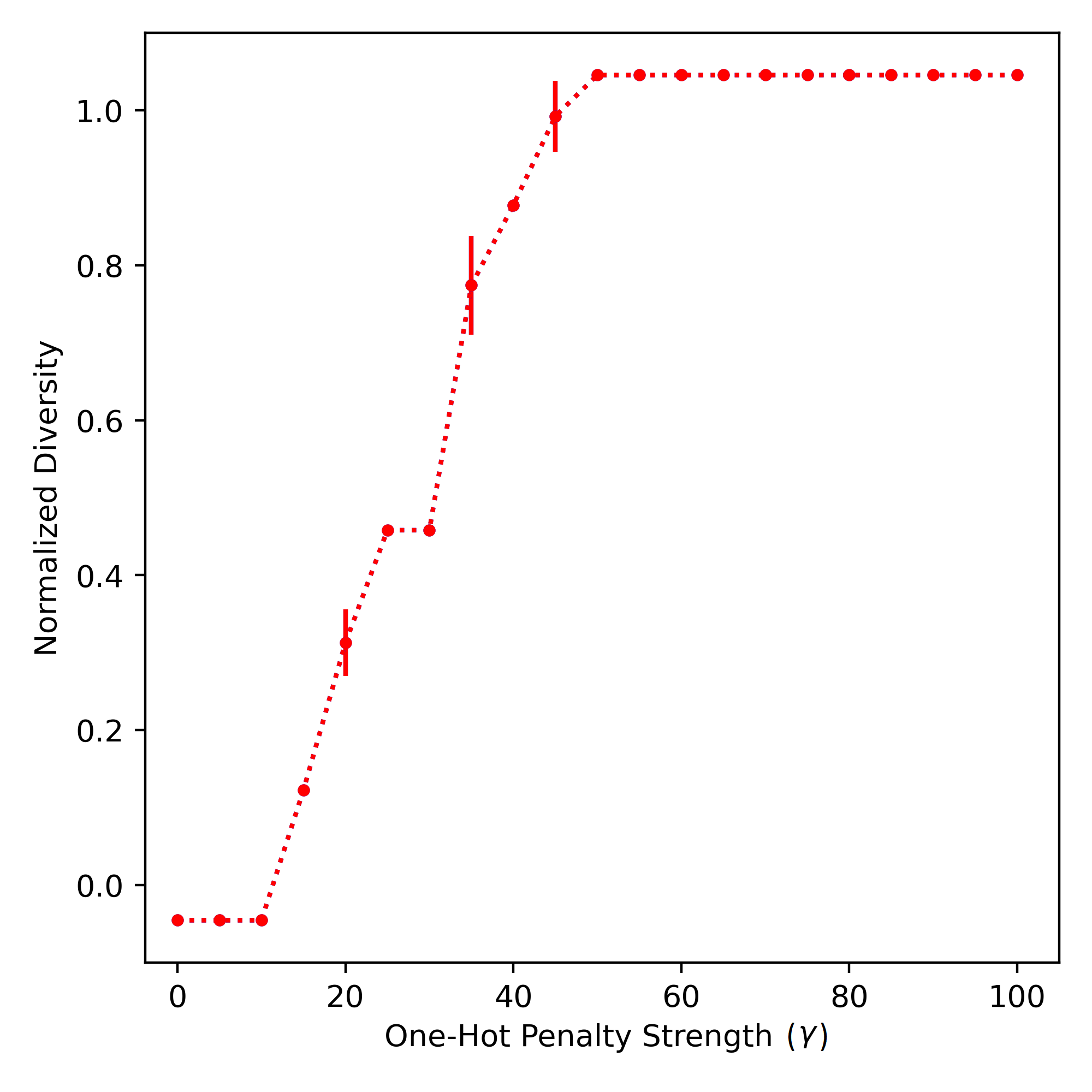}
    \caption{Average normalized diversity returned by D-Wave Advantage 4.1 of one-hot three helix bundle QUBO as a function of one-hot penalty strength (left panel). Each point registers the average of $10^4/n$ replicates of $n$ samples. Highest diversity is achieved for $\gamma\geq50$ with greedy post-processing, and zero diversity for $\gamma\leq10$, where all solutions are taken to non-physical states. The inset shows non-greedy average normalized diversity as a function of penalty strength, with 30 replicates of $10^4$ samples per point. Diversity peaks when $\gamma = 10$. The right panel shows average normalized diversity returned by the QUBO simulated annealing sampler over 30 replicates of $10^4$ samples as a function of one-hot penalty strength. For $\gamma\geq50$, full diversity is achieved.}
    \label{fig:diversity_vs_gamma}
\end{figure*}

%\newpage
\section{Solution quality}

\begin{figure*}[ht!]
    \includegraphics[width=7cm]{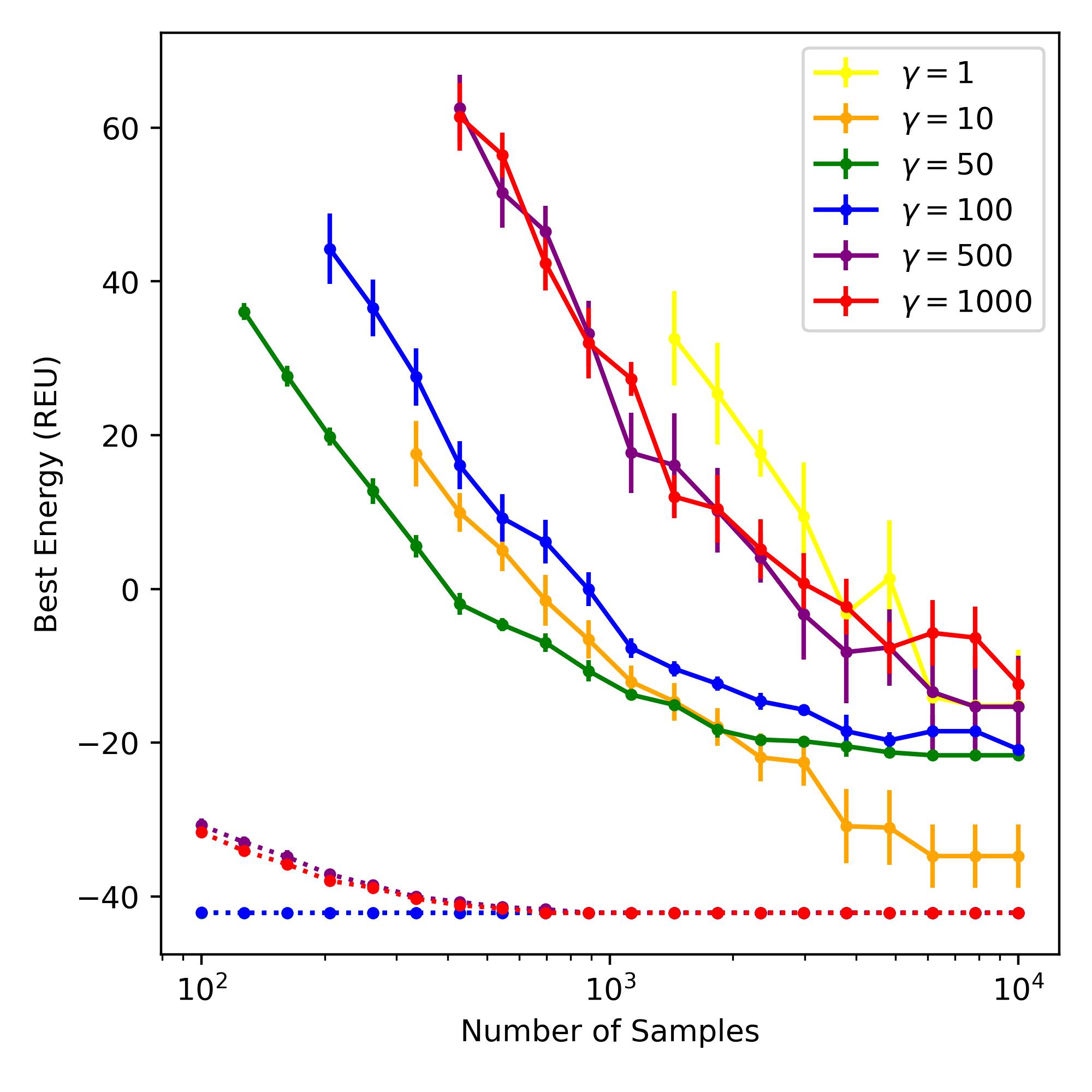}
    \includegraphics[width=7cm]{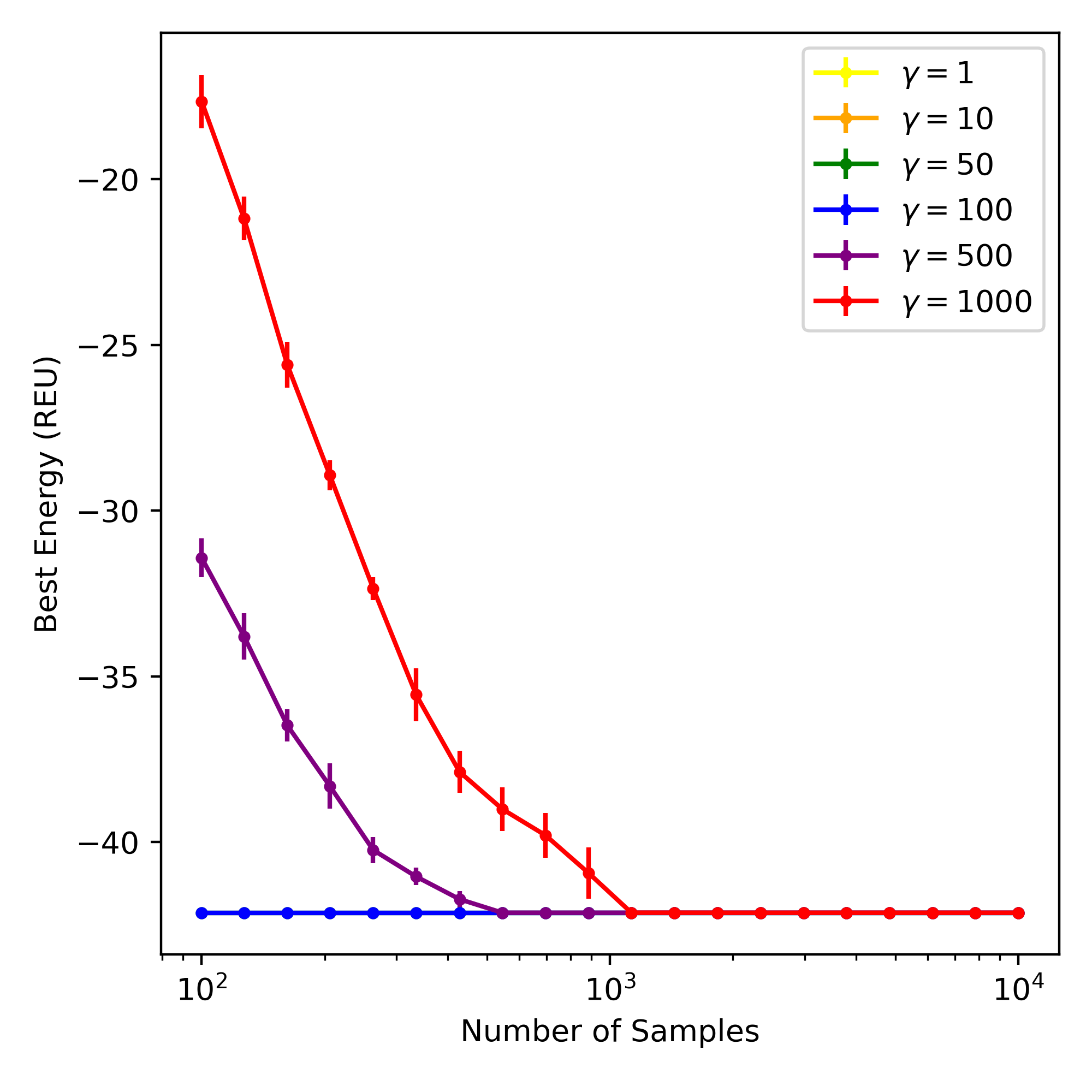}
    \caption{Best energy returned by D-Wave Advantage 4.1 of one-hot three helix bundle QUBO as a function of number of samples (left panel). Each point registers the mean of $25\cdot10^4/n$ replicates, where $n$ is the number of samples. Solid curves correspond to diversity prior to steepest descent post-processing (NG for non-greedy); dotted curves correspond to diversity with steepest descent post-processing (G for greedy). Non-greedy curves show better best energy for $10\leq\gamma\leq100$, while greedy curves show improved best energy for $\gamma\geq100$. Greedy post-processing when $\gamma\leq10$ takes all solutions to non-physical states (zero diversity, no best energy). The right panel shows best energy found by the QUBO simulated annealing solver. For $\gamma\leq10$ all solutions are non-physical; $50\leq\gamma\leq100$ shows best performance.}
    \label{fig:optimum_vs_samples}
\end{figure*}

\end{document}